# ARTICLE

# LAW PROOFING THE FUTURE

Gregory M. Dickinson*

Abstract

*Lawmakers today face continuous calls to "future proof" the legal system against generative artificial intelligence, algorithmic decision-making, targeted advertising, and all manner of emerging technologies. This Article takes a contrarian stance: it is not the law that needs bolstering for the future, but the future that needs protection from the law. From the printing press and the elevator to ChatGPT and online deep fakes, the recurring historical pattern is familiar. Technological breakthroughs provoke wonder, then fear, then legislation. The resulting legal regimes entrench incumbents, suppress experimentation, and displace long-standing legal principles with bespoke but brittle rules. Drawing from history, economics, political science, and legal theory, this Article argues that the most powerful tools for governing technological change—the general-purpose tools of the common law—are in fact already on the books, long predating the technologies they are now called upon to govern, and ready also for whatever the future holds in store.*

*Rather than proposing any new statute or regulatory initiative, this Article offers something far rarer, a defense of doing less. It shows how the law's virtues—generality, stability, and adaptability—are best preserved not through prophylactic regulation, but through accretional judicial decision-making. The epistemic limits that make technological forecasting so unreliable and the hidden costs of early legislative intervention, including biased governmental enforcement and regulatory capture, mean that however fast technology may move, the law must not chase it. The case for legal restraint is thus not a defense of the status quo, but a call to preserve the conditions of freedom and equal justice under which both law and technology can evolve.*

---

* Assistant Professor of Law and, by courtesy, Computer Science, University of Nebraska; Nonresident Fellow, Stanford Law School, Program in Law, Science & Technology; Murry & Polly Bowden Nonresident Fellow, University of Texas at Austin School of Law, Bech-Loughlin First Amendment Center; J.D., Harvard Law School. This Article expands on remarks prepared for the Regulation of Algorithms discussion at the 2025 AALS Annual Meeting on January 10, 2025, in San Francisco, California. Thanks for their comments to Dhruva Krishna, Christina Mulligan, Saurabh Vishnubhakat, Eugene Volokh and participants of the Central States Law Schools Association (CSLSA) Conference and the International Center for Law & Economics (ICLE) Law & Economics Fellows Meeting in October 2025. Thanks also to Linnea Jorgenson and Hannah-Kate Kinney for their research assistance.

## I. Introduction

Daily we hear calls to "future proof the law"[1] against this or that coming danger.[2] The phrase strikes the mind as wise and responsible—an expression of fidelity to progress and foresight. Yet, on reflection, the phrase's implication is curious: that the law must be braced to preserve it against the future's onslaught. This framing places law above the future, as if it is our duty to protect the law and shelter it against the damage of societal change. In truth, the relationship runs in the other direction. It is not the law that must withstand the future, but the future that must be preserved from the threat of law.[3]

---

[1] Typically, such "future proofing" is to be accomplished by enacting a new statute or regulation designed to target some specific technological threat. Examples are beyond enumeration, but the number is hundreds per year even when counting only federal proposals. *See Artificial Intelligence Legislation Tracker*, Brennan Ctr. For Just. (June 30, 2025), https://www.brennancenter.org/our-work/research-reports/artificial-intelligence-legislation-tracker [https://perma.cc/SES6-E4FF] (listing over 150 federal bills concerning AI proposed during the 118th Congress).

[2] Even among legal academics, not typically thought to rank among the world's technical elite, there has been keen interest in hosting workshops, conferences, roundtables, conventions, summits, and every other sort of meeting the mind can invent, all devoted to charting the law's path through technological change. One symposium even bore the "future-proofing" name, Symposium, *Future-Proofing Law: From rDNA to Robots*, 51 U.C. Davis L. Rev. 1 (2017). Recent examples include: *Developments in the Law — Artificial Intelligence*, 138 Harv. L. Rev. 1554 (2025); *Old Law, New Tech: Legal Responses to Emerging Technologies*, 133 Yale L. J. (2024); Symposium, *How AI Will Change the Law*, U. Chi. L. Rev. Online (2025); Symposium, *Common Law for the Age of AI*, 119 Colum. L. Rev. 7 (2019).

This Article challenges the conventional wisdom that rapid technological change demands fresh, similarly rapid lawmaking. That view—deeply intuitive, widely held, and often wrong—treats technological development as a problem to be solved through prescriptive legislation and regulation. In times of innovation, we look to lawmakers for reassurance and intervention against new dangers. We tell ourselves that new tools raise novel dangers and that law must evolve to meet them. And so, from the printing press to social media and generative AI, we have called reflexively for new statutes, new rules, and new agencies.[4] Sometimes, this is sensible. But more often, the instinct to legislate outruns its justification.[5]

The larger danger lies not in what technology might do, but in what we might do in response. When we rush to govern the new, we may inadvertently discard legal virtues so longstanding that they are overlooked—the law's durability and generality, its applicability across time and circumstances to all persons, politically favored or not.[6] As this Article argues, general-purpose law—our long-standing principles providing tort redress for harms, freedom to contract, due process protections for person and property, and criminal sanctions for lawbreakers—stands as ready to govern today's technological change as it was to govern yesterday's.[7] To make that point, the Article employs tools from a variety of disciplines, from history and legal theory to political science, economics, and sociology. The Article begins by placing the modern impulse toward legal intervention in a broader historical pattern: the social cycle that follows in the wake of transformative technologies. We marvel, then we fear, then we regulate. From Gutenberg's press to Facebook's newsfeed and ChatGPT's hallucinatory utterances, we observe the same arc— an initial burst of optimism followed by a public backlash, often culminating in legal restrictions aimed at controlling the unpredictable. Stanley Cohen's theory of moral panic offers one sociological account of this pattern,[8] one for which the historical record—for example, novels, comic books, film, radio, cable, and video games—supplies abundant examples.[9] Legal responses to technology often resemble reactions to vice: panicked, preachy, and prone to overreach.[10]

---

[3] *Cf.* Cass R. Sunstein, Laws of Fear: Beyond the Precautionary Principle 1–9 (2005) (discussing lawmakers' tendency to legislate out of fear and noting that "[a]s a result, the law can be led in unfortunate and even dangerous directions").

[4] Whether and how regulatory mechanisms should be adjusted to accommodate the digital age has spawned a massive literature. For leading recent work, see, e.g., Ganesh Sitaraman, *The Regulation of Foreign Platforms*, 74 Stan. L. Rev. 1073, 1076–83 (2022) (regulation must evolve to reflect platforms' global architectures to avoid ineffective governance); Anu Bradford, *The False Choice Between Digital Regulation and Innovation*, 119 Nw. U. L. Rev. 377, 378–401 (2024) (characterizing the widely supposed tradeoffs between regulation and innovation as a "false choice" and advocating for a balanced regulatory posture); Ryan Calo, *Artificial Intelligence Policy: A Primer and Roadmap*, 51 U.C. Davis L. Rev. 399, 410–30 (2017) (overview of contemporary AI policy debates).

[5] *See* Adam Thierer, Permissionless Innovation 1–4, 29–35 (2014) (explaining how the "precautionary mindset" according to which "public policy is guided at every turn by fear of hypothetical worst-case scenarios" renders innovation less likely and offering proposed data-privacy restrictions as a contemporary example); Maureen K. Ohlhausen, *The Procrustean Problem with Prescriptive Regulation*, 23 CommLaw Conspectus 1, 3 (2014) (FTC Commissioner calling for "regulatory humility" and noting that "[t]he massive benefits of perhaps the most influential technology in history, the Internet, . . . have been a result of entrepreneurs' freedom to experiment").

[6] *See* Richard A. Epstein, *The Static Conception of the Common Law*, 9 J. Legal Stud. 253, 253–56 (1980) (explaining that to provide justice and perform its essential stabilizing function, the law must remain constant even in the face of societal change).

[7] *See generally* Richard A. Epstein, Simple Rules for a Complex World ix–xiv, 53–111 (1995) (advocating a regime that "embraces private property and freedom of contract," and redress for their wrongful deprivation, as "the only one that in practice can offer permanence and stability").

The Article then turns to explore the insights that public choice theory and Austrian economics can offer for technology governance. A key insight is that legislative and administrative processes that produce laws to govern fast-paced technology are themselves neither neutral nor nimble. They are slow-moving, interest-group-driven systems prone to capture[11] by their most organized constituencies—in this context, Big Tech. The concentrated benefits and diffuse costs of poorly designed technology restrictions distort policy toward industry incumbents.[12] In the tech context, this means that the Googles and Facebooks of the world are likely to shape the very rules meant to constrain them.[13]

---

[8] *See generally* Stanley Cohen, Folk Devils and Moral Panics vi–xliv, 1–20 (3d ed. 2002), discussed *infra* Part II.

[9] *See generally* Peter Swirski, *Popular and Highbrow Literature: A Comparative View*, 1 CLCWeb: Compar. Literature & Culture (1999) (discussing the social status and role of these and other forms of popular literature).

[10] For examples and discussion of legal and industry responses, see Bobby Chesney & Danielle Citron, *Deep Fakes: A Looming Challenge for Privacy, Democracy, and National Security*, 107 Calif. L. Rev. 1753, 1771–91 (2019) (fears of deep fakes driving regulatory pressure given potential for misinformation, fraud, and other harms); Kate Klonick, *The New Governors: The People, Rules, and Processes Governing Online Speech*, 131 Harv. L. Rev. 1598, 1622–25 (2018) (voluntary social media censorship undertaken in part to forestall governmental regulation); Jack M. Balkin, *The Three Laws of Robotics in the Age of Big Data*, 78 Ohio St. L.J. 1217, 1217–22 (2017) (discussing the fear of robot insurrection and the famous "laws of robotics" intended to limit that danger).

[11] For the seminal work on this point, see George J. Stigler, *The Theory of Economic Regulation*, 2 Bell J. Econ. & Mgmt. Sci. 3, 3 (1971) (famously concluding that, in general, "regulation is acquired by the industry and designed and operated primarily for its benefit").

[12] *See* Mancur Olson, The Logic of Collective Action 1–5, 11–16 (1965) (diffuse interests of the voting public will fail to motivate voters sufficiently to compete with the narrow interests of a regulated industry).

[13] *See* Julie E. Cohen, *The Regulatory State in the Information Age*, 17 Theoretical Inquiries L. 369, 370–75 (2016) (explaining that new regulatory models have "provided new points of entry for power" and "have tended to be both opaque to external observation and highly prone to capture"); *see also* Gary E. Merchant, *Governance of Emerging Technologies as a Wicked Problem*, 73 Vand. L. Rev. 1861, 1861–63 (2020) (explaining that "[e]xisting regulatory agencies lack the legal authority, expertise, and resources to regulate any of the emerging technologies comprehensively, even if they wanted to," which is one reason they must turn to experts in the regulated industry for guidance).

Yet by acting prematurely, legislatures often commit themselves to failure. Legislatures and regulatory agencies cannot possibly gather the vast, localized, dynamic information needed to make fine-grained, welfare-maximizing decisions.[14] That limitation is all the more acute in fast-moving technological domains such as AI systems.[15] Regulators must guess at the pace and path of innovation, codifying answers before the questions have even stabilized.[16]

Next, the Article examines the danger of legal "ossification"—the tendency of administrative rulemaking to become so procedurally encumbered that agencies are unable or unwilling to update rules in light of new circumstances, and the law becomes outdated.[17] The ossification challenge is especially difficult in the context of quickly evolving technologies. As rulemaking has over the years become procedurally complex, legally risky, and politically fraught, agencies increasingly avoid rulemaking altogether.[18] This has led not to a more responsive legal order, but to one in which guidance documents and informal enforcement actions displace deliberative lawmaking—a result that reduces transparency, undermines legitimacy, and insulates regulators from democratic accountability.[19] Ironically, ossifications means that efforts to tailor rules to technology can produce a legal environment that is both less adaptable and less coherent.

---

[14] *See* F.A. Hayek, *The Use of Knowledge in Society*, 35 Am. Econ. Rev. 519, 521 (1945) (welfare-maximizing legislation requires, beyond scientific knowledge, "knowledge of the particular circumstances of time and place" that are unknown and many times unknowable to lawmakers).

[15] *See* Ohlhausen, *supra* note 5, at 3 (FTC Commissioner Ohlhausen commenting that "[b]ecause it is so difficult to predict the future of technology, government officials, myself included, must approach new technologies and new business models with a significant dose of regulatory humility").

[16] *See* Joseph J. Avery, Patricia Sánchez Abril & Alissa del Riego, *ChatGPT, Esq.: Recasting Unauthorized Practice of Law in the Era of Generative AI*, 26 Yale J.L. & Tech. 64, 89–92 (2023) (AI's strain on regulation of legal professionals); *see also* Cheryl B. Preston, *Lawyers' Abuse of Technology*, 103 Cornell L. Rev. 879, 880 (2018) (proposals to update professional responsibility principles for lawyers to address technology developments); Andrew D. Selbst, *An Institutional View of Algorithmic Impact Assessments*, 35 Harv. J.L. & Tech. 117, 122–25 (2021) (challenges to proposed algorithmic impact assessments); Ignacio N. Cofone, *Servers and Waiters: What Matters in the Law of A.I.*, 21 Stan. Tech. L. Rev. 167, 169–70 (2018) (challenges of placing AI agents and robots within existing legal structures); Benjamin G. Edelman & Damien Geradin, *Efficiencies and Regulatory Shortcuts: How Should We Regulate Companies Like Airbnb and Uber?*, 19 Stan. Tech. L. Rev. 293, 294–95 (2016) (difficulties in fitting new business models within existing legal categories).

[17] *See* Thomas O. McGarity, *Some Thoughts on "Deossifying" the Rulemaking Process*, 41 Duke L.J. 1385, 1385–86 (1992) (describing how the accumulation of procedural and analytical mandates has slowed rulemaking and discouraged agencies from undertaking regulatory updates).

[18] *See id.* at 1386 (characterizing the rulemaking process as "heavily laden with additional procedures, analytical requirements, and external review mechanisms").

[19] *See id.* at 1442 (reasoning that "[a] wholesale shift to these less formal devices could leave regulatory agencies much less accountable to the public and could pave the way to arbitrary decisionmaking"). For recent exploration of these themes in the age of AI, see David F. Engstrom & Daniel E. Ho, *Algorithmic Accountability in the Administrative State*, 37 Yale J. Reg. 800, 804–07 (2020) (agency use of algorithms in the work of governance and a proposal for monitoring and oversight); *see also* Cristina I. Ceballos, David F. Engstrom & Daniel E. Ho, *Disparate Limbo: How Administrative Law Erased Antidiscrimination*, 131 Yale L.J. 370, 384, 399–401 (2021) (bias concerns in agency use of digital-governance tools).

Finally, even after considering the many challenges to technology governance, the Article concludes on a note of optimism. The point is not that law should never respond to new technology, but that a well-designed, time-tested system like our own rarely needs to. Legal change in response to even the most innovative technologies should be rare, modest, and general.[20] Just as well-written computer code is reusable, good law will apply across cases, actors, and epochs. Over time, the best legal systems are those that resist the temptation to engineer against every contingency. The law draws strength from abstraction, stability, and resistance to fads.[21]

This Article thus takes the path less followed in adopting a skeptical view of techno-legal exceptionalism. It suggests that the core problems posed by artificial intelligence, digital platforms, biotechnology, and other frontier technologies are not so different from those that the law has long confronted as to require radical reengineering.[22] Fraud is still fraud. Harm is still harm. Rights, duties, and remedies still do their quiet work. In the end, it is not the law that is unprepared for technology but current policy discourse that is unprepared to offer legal stability.[23]

The stakes are high. As societies increasingly govern through statute and regulation, the costs of legal error—especially legal overreaction—grow larger.[24] Premature regulation may lock in early designs,[25] stifle experimentation,[26] and entrench incumbents.[27] Over time, this can channel innovation away from contested domains and into less regulated spaces,[28] not because those areas are more fruitful, but because they are more legally tolerable.

---

[20] *See infra* Part III. Other recent work taking this general approach includes Orin S. Kerr, *Norms of Computer Trespass*, 116 COLUM. L. REV. 1143, 1146–47, 1153–73 (2016) (developing a common-law approach to interpreting the Computer Fraud and Abuse Act's trespass-like provisions and suggesting legal clarity in digital trespass will emerge from societal norms); *see also* Anat Lior, *Insuring AI: The Role of Insurance in Artificial Intelligence Regulation*, 35 HARV. J.L. & TECH. 467, 524–27 (2022) (setting forth a regime of private insurance as an alternative in some contexts to regulation); Pamela Samuelson, *Functionality and Expression in Computer Programs: Refining the Tests for Software Copyright Infringement*, 31 BERKELEY TECH. L.J. 1215, 1223 (2016) (suggesting that general doctrinal rules of copyright infringement can be positioned to support ongoing innovation); EPSTEIN, *supra* note 7, at ix–xiv, 327–31 (illustrating the point with the law of contracts, explaining that the "law of contract is able to keep up with the most rapid-fire innovations, as long as it remains uncluttered with restraints" and that "we could [even] do as well with the Roman law of contract" with only "a little refurbishing at the edges").

[21] *See* F. A. HAYEK, LAW, LEGISLATION, AND LIBERTY 28–33 (1973) (urging generally applicable, abstract rules, concluding that it is "[t]he *hubris* of reason" to believe that one "can dispense with abstraction and achieve full mastery of the concrete").

[22] *Cf. id.* at 46–50 ("[R]ules governing a spontaneous order must be independent of purpose and be the same . . . for whole classes[.] They must . . . be rules applicable to an unknown and indeterminable number of persons and instances.").

[23] *See* Epstein, *supra* note 6, at 255 (observing lawmakers' proclivity for legal innovation, but noting that "there are many areas in which the legal system need not take into account changes in either social behavior or technological patterns").

[24] *See* THIERER, *supra* note 5, at 82 (fear-induced regulations will sacrifice our "freedom to experiment" and "paralyze[] our innovative spirit").

[25] *See infra* Part IV.E; Bradford, *supra* note 4, at 415–17 (discussing how poorly timed or overhasty regulation can "lock in" early technological designs and offering potential strategies for avoiding that result).

[26] *See* Jon Truby, Rafael D. Brown, Imad A. Ibrahim & Oriol C. Parellada, *A Sandbox Approach to Regulating High-Risk Artificial Intelligence Applications*, 13 EUR. J. RISK REGUL.

This Article proceeds in three Parts. Part II traces the historical pattern of legal overreaction to new technologies, showing how initial enthusiasm often gives way to social panic and legal suppression. It draws on examples from print, broadcasting, video games, and the internet, illustrating the cyclical nature of technological scares and the legal responses they generate. Part III offers a defense of general-purpose law as well suited to manage technological change, emphasizing the virtues of abstraction, neutrality, and organic legal development. Part IV analyzes the institutional pathologies that lead to poorly designed technology law, drawing on public choice theory, the Hayekian knowledge problem, and the phenomenon of legal ossification. The Article concludes with a call for legal restraint—not as a rejection of technology regulation but as a higher form of it.

## II. The Human Tendency Toward Fresh Lawmaking

The natural, perhaps inevitable response to new technology is first to marvel, briefly to hope for what might come, and ever after to fear the loss that change may bring. That is, until what was the shiny new becomes the old familiar and the cycle starts afresh. As the fear sets in, we throw into gear the lawmaking process to regulate the unknown. It is virtually impossible to ignore the impulse; it is what one's gut commands and exactly how one will proceed unless the conscious mind can persuade otherwise. With that, this Article aims to help.

---

270, 273–77 (2021) (observing that early regulation risks undermining innovation and proposing regulatory sandboxes as a more flexible alternative); *see also infra* note 187 (collecting sources discussing and estimating effects of GDPR's data-sharing restrictions on innovation in app design).

[27] *See infra* Part IV.C. One recent example is Mississippi's age-verification law, which requires social media platforms to verify that their users are eighteen years of age. *Walker Montgomery Protecting Children Online Act*, 2024 Miss. Laws ch. 456 (codified at Miss. Code Ann. §§ 45-38-1 to -13). Small competitors to the major platforms, such as Bluesky and Mastodon, have found compliance impossible or prohibitively expensive. Drew Harwell, *'Scan Your Face' Laws For The Web Are Having Unexpected Consequences*, Wash. Post (Aug. 31, 2025), https://www.washingtonpost.com/technology/2025/08/31/age-verification-uk-porn-sites/ [https://perma.cc/5242-DZP7] (noting Bluesky's decision to discontinue operations in the state); Sarah Perez, *Mastodon Says It Doesn't 'Have the Means' to Comply with Age Verification Laws*, Tech-Crunch (Aug. 29, 2025), https://techcrunch.com/2025/08/29/mastodon-says-it-doesnt-have-the-means-to-comply-with-age-verification-laws/ [https://perma.cc/3MP9-TSAK].

[28] Movement may also be geographical. For discussion of some of the challenges this poses for lawmakers, see Ganesh Sitaraman, *The Regulation of Foreign Platforms*, 74 Stan. L. Rev. 1073, 1135–52 (2022) (discussing the TikTok and WeChat examples through historical lens); *see also* Jennifer Daskal, *Microsoft Ireland, the Cloud Act, and International Lawmaking 2.0*, 71 Stan. L. Rev. Online 9 (2018) (discussing complexity of domestic and international law regarding requests for data held in foreign jurisdictions).

### A. *Gutenberg and the Machine That Launched a Thousand Laws*[29]

History shows the pattern repeating again and again: new technologies elicit initial marvel, followed by apprehension regarding potential societal change or ideological shifts, which in turn galvanizes lawmaking bodies to restrict the new and unknown. The process is a natural, almost inevitable human impulse.

Consider the reception of the printing press in Europe. Initially, the Roman Catholic Church viewed the invention as a valuable instrument for propagating the faith, even hailing it as a "divine art."[30] The Church's stance quickly shifted, however, as the Protestant Reformation gained momentum during the fifteenth and sixteenth centuries, and it became clear that the press could be put to many purposes beyond spreading orthodox doctrine. Indeed, Martin Luther's Ninety-five Theses, famously affixed to the church door in Wittenberg in 1517, would likely have remained a localized protest without the amplifying power of the printing press.[31] Instead, the document was rapidly reproduced and widely distributed, igniting a theological debate that fundamentally challenged the Church's authority. The press thus emerged as a potent tool for dissenting voices, enabling religious reformers to circumvent traditional channels of control and circulate their ideas, Bible translations, and critiques of the ecclesiastical order directly to a burgeoning reading public.[32] Demonstrating the feared power of the printed book, the Church aimed to stop the presses, even coining the new sin of "bibliolatry"—improper reverence for books so excessive as to interfere with one's worship of God.[33] The widespread dissemination of heterodox teachings that the printing press enabled, perceived as dangerous to both clerical power and to the spiritual welfare of the populace, prompted Church officials to impose numerous restrictions,[34] including a requirement of papal authorization for new publications.[35]

---

[29] *Cf.* Christopher Marlowe, The Tragical History of the Life and Death of Doctor Faustus, act V, scene i, lines 99–100 at 92 (John D. Jump ed., Harvard University Press 1962) (1604) ("Was this the face that launch'd a thousand ships / and burnt the topless towers of Ilium?").

[30] Jacob Mchangama, Free Speech: A History from Socrates to Social Media 66 (2022) (noting the Church's initial embrace of the printing press before fearing its potential to undermine authority); Eisenstein, The Printing Press as an Agent of Change 118, 317 (1979) (observing that early Catholic authorities, far from opposing Gutenberg's press, often embraced it as a tool for defending orthodoxy and disseminating approved works); *see also* Zack Kertcher & Ainat N. Margalit, *Challenges to Authority, Burden of Legitimization: The Printing Press and the Internet*, 8 Yale J.L. & Tech. 1, 15–16 (2006) (same).

[31] *See* Mchangama, *supra* note 30, at 67–74 (describing how Luther's Ninety-five Theses, quickly reproduced in pamphlets and vernacular translations, spread "like wildfire" across Europe).

[32] *See id.* at 67 (noting that the press allowed reformers to bypass church gatekeepers and disseminate their works to lay readership).

[33] Eisenstein, *supra* note 30, at 326 (discussing Catholic authorities' warning against "bibliolatry").

One direct response to this perceived threat was the infamous *Index Librorum Prohibitorum*, the Index of Prohibited Books—a comprehensive catalogue of publications deemed heretical or morally perilous, the reading of which by Catholics was strictly forbidden.[36] Ostensibly aimed at preserving public welfare, these measures sought to control the intellectual foundations of human society. An invention with immense power and potential thus quickly became entangled in a web of restrictions designed to safeguard existing authorities and, purportedly, protect the populace from itself. The Church's initial embrace morphed into fear, producing a heavy regulatory response to what it could no longer control.

A similar story played out within secular governments, especially in England,[37] where concerns gravitated less toward religious dogma and more towards criticism of the monarch. The Crown, much like the Church, initially recognized the press's usefulness for disseminating laws, proclamations, and official news.[38] Yet, political dissent, mirroring religious reform, also found a powerful new amplifier in printed materials.[39] Critiques of royal policy, calls for political reform, and even arguments for outright sedition could now be replicated and distributed with a speed and breadth unimaginable in the era of handwritten manuscripts. This presented an immediate threat to the realm's stability and the monarch's absolute authority.

As with the Church, the danger of the printing press was, of course, not expressly framed to the public merely as a risk to the King's power. Instead, restrictions on printing were rationalized as being for the public good, aimed at preventing unrest and preserving governmental stability. The prevailing argument posited that unbridled speech, especially printed matter, could incite rebellion, corrupt public morals, or disseminate misinformation leading to societal chaos. For instance, the English Crown routinely targeted seditious and libelous statements to suppress any printed material criticizing the government or promoting ideas contrary to royal authority.[40]

---

[34] *See* MCHANGAMA, *supra* note 30, at 66–67 (describing Church's efforts to curb influence of printed works through censorship and other controls).

[35] *See id.* at 67 (noting Church imposing rule that no book could be published without prior approval); EDOARDO TORTAROLO, THE INVENTION OF FREE PRESS: WRITERS AND CENSORSHIP IN EIGHTEENTH CENTURY EUROPE 6–7 (2016) (discussing pre-publication controls to prevent dissemination of texts deemed threatening to political or religious order).

[36] *See* EISENSTEIN, *supra* note 30, at 347–49, 355, 411 (discussing the *Index Librorum Prohibitorum*, or "Index").

[37] Efforts to censor and regulate publication went even further in the Ottoman Empire. Arabic characters were not added to the printing press until the eighteenth century. Even after the introduction of Arabic characters, the publication of Islamic texts was prohibited until the nineteenth century. For the history of the printing press in the Islamic world, see generally JOHANNES PEDERSEN, THE ARABIC BOOK 131–41 (Robert Hillenbrand ed., Geoffrey French trans., Princeton Univ. Press 1984); LUCIEN FEBVRE & HENRI-JEAN MARTIN, THE COMING OF THE BOOK: THE IMPACT OF PRINTING 1450–1800 212–15 (Geoffrey Nowell-Smith & David Wootton eds., David Gerard trans., NLB 1976) (1958).

[38] *Cf.* EISENSTEIN, *supra* note 30, at 104–05 (discussing use of the press to publish governmental documents).

[39] *See* MCHANGAMA, *supra* note 30, at 75–77 (explaining that controversial materials were not a new phenomenon, but could be disseminated more quickly with the advent of the printing press).

"Extreme times" called for extreme measures, or so the argument went. The Crown's traditional punitive tools (including burning at the stake) proved insufficient on their own, given the terrific quantities of information that the printing press enabled. The sheer volume of printed material rendered post-publication punishment insufficient as a deterrent; by the time a seditious pamphlet could be identified and its authors or printers apprehended, countless copies might already be in circulation, having already done their ostensible damage.[41] Consequently, the government shifted its approach to proactive control.[42]

Post-publication physical punishment for content deemed treasonous or heretical was augmented by a system of prior restraint, which criminalized publication without explicit Royal preapproval.[43] This system entailed licensing printers, requiring them to secure official permission to operate, and establishing censors responsible for reviewing material prior to its release.[44] During this time the Star Chamber, the infamous English royal court, became a formidable instrument of censorship, issuing decrees that controlled printing, limited the number of authorized printers, and mandated licensing for all books.[45] Violations could lead to severe penalties for printers and authors alike, including imprisonment, substantial fines, and physical mutilation.[46]

Importantly, these measures were not merely extensions of existing laws designed to maintain order; they represented a profound shift to control thought and expression, in direct response to a new technology that had revolutionized the dissemination of ideas. The fear by those in power was palpable, as an unregulated press held the potential to dismantle the existing social and political order. Both Church and State, facing a rapidly evolving information landscape, reacted naturally by regulating the unknown, often framing restrictions as essential for public welfare, even if their underlying, sometimes subconscious, motivation was preservation of established power.

---

[40] *See id.* at 76–79 (discussing criminalization of seditious statements and misinformation that undermined respect for the Crown).

[41] *See id.* at 76 (observing that speed and scale of print production made it difficult for authorities to suppress offending works before they had been widely distributed and read).

[42] *See id.* at 76–77 (noting how difficulty halting the spread of subversive works after printing led authorities to adopt pre-publication censorship and indices of forbidden literature).

[43] *See id.* at 77, 112 (recounting how English law supplemented harsh post-publication penalties with licensing requirements).

[44] *See id.* at 77, 112, 115 (explaining how English licensing regime both restricted the number of authorized printers and titles and empowered designated censors to vet works).

[45] *See id.* at 104 (providing a brief history of the Star Chamber's controls over the press). For a detailed discussion of the Star Chamber's checkered history, see generally STAR CHAMBER MATTERS: AN EARLY MODERN COURT AND ITS RECORDS (K. J. Kesselring & Natalie Mears eds., Univ. of London Press 2021); J.A. GUY, THE CARDINAL'S COURT: THE IMPACT OF THOMAS WOLSEY IN STAR CHAMBER (1977). For a somewhat more sympathetic perspective, see Thomas G. Barnes, *Star Chamber Mythology*, 5 AM. J. LEGAL HIST. 1 (1961).

[46] *See* MCHANGAMA, *supra* note 30, at 104.

### B. Magazines and Brain-Rot Contagion

Somewhat closer to the present, consider also the later alarm over disposable print media. One delightful example comes from an article titled "Lively Comment Upon Dewey, the Newest Fads in Publishing and English Mass Meetings: Brain-Rot Contagion" published in 1899 in the *Brooklyn Daily Eagle*, which warned of a "brain-rot contagion" induced by the widespread circulation of magazines:[47]

> The number of people who think like birds, in little broken thoughts, will be greatly enlarged. *Millions upon millions* of American boys and girls and men and women will be, like their English cousins, unable to learn anything, to know anything well and to concentrate their minds upon anything.

This fear sounds distant, yet familiar. Cyclical apprehension of whatever media is new in the moment is not coincidental; it can be explained as a recurring sociological phenomenon known as a "moral panic."[48] The phrase, coined by Stanley Cohen, describes a period of intense public fear and anxiety over a perceived threat to societal values or norms, which is often amplified by media representations and leads to calls for stricter control and regulation.[49] These panics often target specific groups or cultural forms, portraying them as "folk devils" that embody the perceived threat.[50] The arguments asserted in support of societal counteractions predictably frame the new or unfamiliar as inherently dangerous before calling for its restriction.[51]

Beyond the "brain-rot" attributed to magazines, historical anxieties have repeatedly coalesced around other forms of popular print media, deemed detrimental to public morals, intellectual rigor, or social order. The novel, for instance, was a significant target of such cultural opprobrium, particularly in Britain during the late eighteenth and early nineteenth centuries.[52]

---

[47] Julian Ralph, Editorial, *Lively Comment Upon Dewey, the Newest Fads in Publishing and English Mass Meetings: Brain-Rot Contagion*, BROOK. DAILY EAGLE, Oct. 8, 1899, at 15.

[48] *See* COHEN, *supra* note 8, at xix–xx, 1–4 (describing how societies are recurrently gripped by "moral panics" in which new groups or media are exaggerated as threats to social order).

[49] *See id.* at 1–5 (introducing the concept of "moral panic" as a recurring social phenomenon in which a person, group, or condition is cast as a threat to societal values, often leading to exaggerated public reactions and policy responses).

[50] *See id.* at 1–4 (explaining how groups such as hippies, student militants, vandals, rockers, drug fiends, skinheads, and soccer hooligans come to be cast as symbolic folk devils).

[51] *See id.* at vi–xliv, 89–93 (dangers magnified through media and political discourse and perceived risks used to justify heightened social control).

[52] Gary Kelly, *This Pestiferous Reading: The Social Basis of Reaction Against the Novel in Late Eighteenth- and Early Nineteenth-Century Britain*, 4 MAN AND NATURE 183, 188–93 (1985) (examining attacks on the novel in Britain as reflections of broader class and ideological

Critics condemned novels for fostering idleness, especially among women, who were perceived as susceptible to the genre's seductive narratives, which could distract them from their domestic duties and lead to their moral corruption.[53] It was feared that novels were so immersive (like virtual-reality headsets) that they permitted a dangerous escapism that would promote unrealistic expectations of life and incite social unrest by exposing readers to subversive ideas.[54] The prevailing sentiment was that such "pestiferous reading" undermined the social fabric by promoting emotional indulgence over rational thought and duty.[55]

The division between supposedly corrupting popular literature, such as the common novel, and intellectually uplifting, "highbrow" literature, continues today and has been a perennial feature of media moral panics. Mass-produced, "lowbrow" print materials are scrutinized and condemned because of their accessibility to a broad populace and their perceived lack of artistic or intellectual merit.[56] The driving fear, often unstated, is that such easily digestible content bypasses critical faculties, leading to a passive, unthinking consumption that renders the public susceptible to manipulation or moral decay.[57]

A vivid twentieth-century manifestation of such anxiety occurred in the 1950s with the widespread condemnation of comic books. What began as a burgeoning and largely unregulated industry soon faced intense public and political pressure, fueled by sensationalized media reports and the influential critiques by figures such as psychiatrist Fredric Wertham.[58] Wertham's seminal 1954 book, *Seduction of the Innocent*,[59] posited a direct causal link between reading comic books and juvenile delinquency, violence, and sexual perversion. He argued that comics desensitized children to violence, glamorized crime, and introduced them to inappropriate sexual themes, thereby corrupting their morals and inciting antisocial behavior.[60] The pervasive fear was

---

tensions, with critics portraying fiction, especially works associated with aristocratic culture, as socially and morally corrupting).

[53] *See id.* at 192 (novels associated with "decadent aristocratic culture" and "inspired fantasies of social climbing").

[54] *See id.* at 189 (exploring how critics of the period portrayed novels as dangerously absorbing, capable of shaping readers' values and aspirations and introducing socially disruptive ideas).

[55] *See id.* at 183–93.

[56] *See* Swirski, *supra* note 9, at 2 (surveying sociological data and aesthetic critiques to challenge the entrenched hierarchy between "highbrow" and popular literature, arguing for "greater literary democracy" and criticizing the academic habit of leaving "97% of its subjects camping outside the city gates").

[57] *See id.* at 10 (reciting common critiques of popular literature, including claims that mass-produced fiction fosters escapism and dulls critical engagement and arguing that such charges are often unsupported by empirical evidence and overlook the genre's aesthetic value).

[58] *See* Scott Woodcock, *Fear Within the Frames: Horror Comics and Moral Danger*, 53 Can. J. Philos. 535, 535–36 (2023) (tracing moral uproar over comic books and subsequent industry self-regulation to Fredric Wertham's work).

[59] Fredric Wertham, Seduction of the Innocent (1st ed. 1954).

[60] *See id.* at 118 (arguing that crime comics "create a readiness for temptation" by immersing children in "unwholesome fantasies" and "suggest[ing] criminal or sexually abnormal ideas").

that these colorful, seemingly innocuous publications were actively contributing to a national crisis of youth crime and moral decay.

The ensuing public outcry and congressional hearings, most notably the 1954 Senate Subcommittee on Juvenile Delinquency,[61] mirrored the earlier anxieties surrounding the printing press and novels. Testimonies and media narratives amplified the perceived dangers, creating a moral panic that ultimately forced the comic book industry to self-regulate by establishing the Comics Code Authority.[62] This self-censorship mechanism, while ostensibly voluntary, was agreed to only under threat of coercive governmental regulation.[63] It imposed strict content guidelines that stifled creative expression and dramatically altered the landscape of American comic books for decades.[64] The case of comic books perfectly illustrates the cyclical nature of societal reactions to new media:[65] initial apprehension morphs into a moral panic, which then galvanizes a regulatory response, ostensibly to protect the public from itself, but often with profound implications for freedom of expression and cultural and economic development.

There has been a consistent historical pattern[66] of anxiety about new media, from the early condemnation of novels and "magazine brain rot," to 1950s comic books and television brain rot,[67] and now of Generation Z's[68] even scarier sounding "bed rotting."[69] Across diverse historical episodes, societies consistently react with fear to evolving trends in media consumption, often reaching for the tools of regulation. Restrictions are rationalized as essential for public welfare, even if, in retrospect, history often reveals an underlying motivation to preserve incumbent power and political structures from destabilizing effects of information dissemination.[70] The persistent fear is that readily accessible, popular media formats possess an insidious power to corrupt individuals and unravel the fabric of society, thereby justifying stringent oversight and control.

---

[61] *Juvenile Delinquency (Comic Books): Hearing on S. Res. 190 Before the Subcomm. to Invest. Juv. Delinq. of the S. Comm. on the Judiciary*, 83d Cong. 16, 109 (1954) (a psychiatrist who studied the effects of comic books on children stated "[W]e found that comic-book reading was a distinct influencing factor in the case of every single delinquent or disturbed child we studied," and a Senator inquired of a publisher that "in little over 3½ months you sell more of your crime and horror than you sell of the Bible stories?").

[62] Comics Mag. Ass'n of Am., *Code of the Comics Magazine Association of America*, (Oct. 26, 1954), https://cbldf.org/the-comics-code-of-1954/ [https://perma.cc/J2GQ-B7TQ]. For a recent, retrospective account, see Karen MacPherson, *These Books are Real Books*, Wash. Post, Mar. 1, 2020, at E10 ("The Comics Code required that comic books show respect for established authority and banned nudity and explicit violence. While voluntary, the code was followed by many publishers, resulting in an industry focused largely on producing simple, laugh-inducing kids' comics like 'Archie'").

[63] *See* MacPherson, *supra* note 62 ("Worried that the federal government would try to regulate their industry, comics publishers established the Comics Code Authority as an alternative to government regulation.").

[64] *See* Marc H. Greenberg, Comic Art, Creativity and the Law 2 (2d ed. 2022) (discussing how "the application of law and legal doctrine has worked to constrain th[e] creative process").

[65] For a broader discussion of societal reactions to other forms of new media, see generally Kristine Drotner, *Dangerous Media? Panic Discourses and Dilemmas of Modernity*, 35 Paedogogica Historica 593 (1999); Kirsten Drotner, *Modernity and Media Panics*, *in* Media Cultures: Reappraising Transnational Media (Michael Skovmand & Kim Christian Schrøder eds., Routledge 1992).

[66] The pattern long predates the novel. At least as early as Plato, leading thinkers were concerned that virtue would be undermined by the content of prevailing popular entertainment. *See, e.g.*, Plato, *Laws*, (Trevor J. Saunders trans.) *in* Plato Complete Works 816e–817e, 1483–84 (John M. Cooper & D.S. Hutchinson eds., Hackett Publ'g Co. 1997) (rejecting work of both the comedic and dramatic poets, reasoning that "if we intend to acquire virtue . . . we can't be serious and comic too" and regarding the dramatic poets that legal authorities should "first decide whether [proposed] work is fit to be recited").

### C.  *Purveyors of Elevator Terror*

For another example, consider elevators. Imagine how scary they must have been to their first riders. Even today, with a decades-long safety record, elevators are still common triggers of claustrophobia, acrophobia, agoraphobia, and other fears.[71] Their emergence in the nineteenth century thus quite naturally provoked an initial response of fear in society, followed closely by regulation.

Prior to the mid-nineteenth century, vertical movement within multistory structures was a laborious affair, largely confined to staircases or rudimentary hoisting mechanisms that lacked even basic safety features.[72] The very concept of a machine that could safely transport humans vertically within a building

---

[67] *See, e.g.*, TV DANGER STRESSED: *Competition to Education Seen by Woman Member of F.C.C.*, N.Y. Times, Apr. 19, 1951, at 46 (warning that "[TV] would become a competitor to education unless educators made use of it"); *Speaker Tells Of Dangers of TV Influence*, Wash. Post, May 15, 1954, at 33 (suggesting television may cause juvenile delinquency because of crime shows and lack of religious programming); Willy Wonka & the Chocolate Factory (Paramount Pictures 1971) (depicting danger to young boy, "Mike Teavee," who lacks creativity because of his obsession with television); Charlie and The Chocolate Factory (Warner Bros. Pictures 2005) (remake with character who is obsessed with the internet and video games).

[68] *See* Leo Sands, *U.K. Conservatives want Mandatory National Service. Gen Z is Cringing*, Wash. Post, May 28, 2024, at A9 (discussing bed rotting as "the practice of spending hours in bed during the day—often with snacks or an electronic device"); Megan Marples, *Bed Rotting: TikTok's latest trend reveals toxic side of self-care*, Phil. Trib., July 25, 2023, at 2B. (attributing the phrase's invention to Gen Z).

[69] A phrase with a similar meaning, "hurkle-durkle," has been used in the Scottish dialect for over 200 years. *Hurkle*, Dictionaries of the Scots Language, https://dsl.ac.uk/entry/snd/hurkle_v1_n1 [https://perma.cc/4LVM-F6QP] (defining hurkle-durkle as "to lie in bed or lounge about when one should be up and about").

[70] *See* Drotner, *supra* note 65, at 598–604 (explaining that although historical restrictions on media were couched in moral or educative language, they often masked efforts to stabilize existing political and cultural hierarchies).

[71] *Elevator Phobia*, APA Dictionary of Psychology (last updated Apr. 19, 2018), https://dictionary.apa.org/elevator-phobia [https://perma.cc/XR7T-WK48] (noting that the fear of elevators "may represent fear of heights (acrophobia), fear of being enclosed (claustrophobia), or fear of having panic symptoms (e.g., as occurs with agoraphobia)").

[72] Stephen R. Nichols, *The Evolution of Elevators: Physical-Human Interface, Digital Interaction, and Megatall Buildings*, *in* Frontiers of Engineering: Reports on Leading-Edge Engineering from the 2017 Symposium 85, 85–96 (2018) (surveying the historical

was, at its inception, profoundly counterintuitive. The idea challenged deeply ingrained perceptions of gravity and stability.[73] Early contrivances, primarily utilitarian lifts for freight, were notoriously unreliable, often with catastrophic consequences, which fostered public distrust of any vertical conveyance other than the power of one's own two feet.[74] It was against this backdrop of pervasive skepticism and genuine peril that Elisha Graves Otis introduced his revolutionary "safety hoist."[75]

Otis, originally a master mechanic in a bedstead factory in Yonkers, New York, conceived of a simple yet ingenious device: a spring-loaded ratchet system designed to automatically engage with notched guide rails should the hoisting rope fail, thereby preventing the car from plummeting.[76] His innovation was not the elevator itself, which had existed in various forms for millennia, but rather the assurance of its safety.[77]

Yet, given humanity's general apprehension of change, convincing a wary public that his contraption was not a death trap required more than mere engineering prowess; it demanded the sort of theatrical demonstration typical of the times. The pivotal moment arrived at the 1854 Exhibition of the Industry of All Nations at New York's Crystal Palace.[78] In a spectacle orchestrated with the flair of a showman, Otis, standing confidently on an open elevator platform suspended high above a throng of anxious onlookers, dramatically ordered an assistant to sever the single hoisting rope.[79] Gasps surely rippled through the crowd as the platform began its uncontrolled descent, only for the safety brake to engage, bringing the car to an abrupt and secure halt after a fall of mere inches. "All safe, ladies and gentlemen! All safe!" Otis proclaimed.[80] This audacious display, repeated countless times over subsequent months, helped transform public perception.[81]

---

development of elevator technology from ancient hoists to modern automated systems and examining how those innovations, including in safety, have enabled the creation of modern buildings).

[73] *Id.* at 86–87 (discussing how the introduction of safety mechanisms overcame public skepticism rooted in fears about vertical motion and the risks of mechanical failure).

[74] *Id.* (discussing how public fear of elevators, due to "[f]raying rope and other mechanical failures" being "common causes of dangerous accidents," resulted in elevator use exclusively for cargo).

[75] The leading modern biography of Otis is Joseph Goodwin, Otis: Giving Rise to the Modern City 5–20 (2001) (tracing Otis's development of the safety hoist within the broader context of mid-nineteenth-century American industrialization and how he transformed public perceptions of elevator safety).

[76] Nichols, *supra* note 72, at 86–87 (explaining how Otis's invention of a fail-safe braking mechanism marked a turning point in elevator engineering).

[77] *See id.* at 85–87.

[78] *See id.* at 86–87 (describing the demonstration and how it affected public perception of elevators).

[79] *Id.* at 87 (depicting a contemporary illustration of the event, which was put on by showman P.T. Barnum).

[80] *Id.*

[81] Goodwin, *supra* note 75, at 5–20 (describing how the repeated stunt directly shifted audience confidence and generated commercial demand).

Still, the elevator took time to earn the public's trust. In 1857, the Haughwout department store in New York was the first to install a passenger elevator, which it envisioned its customers would use to travel between the levels of its five-story building.[82] The plan was scrapped, however, and the elevator removed because customers refused to use it.[83] Slowly, however, utility overcame fear, and elevators became widely accepted. Hotels, for example, found their patrons to be somewhat more receptive to the new machines. With the help of elevators, guests found it much easier to transport luggage to their rooms, a benefit that perhaps helped to make them more willing to brave the perceived danger.[84]

Public apprehension remained, however, and efforts came quickly to restrict and regulate elevator use.[85] Despite Otis's compelling demonstration and Haughwout's subsequent installation of the first passenger safety elevator, deep-seated anxieties persisted. The notion of enclosed elevator cars rapidly ascending and descending must have remained unsettling to many, giving rise to popular fears and calls for elevator-specific lawmaking.[86]

The legal response to this nascent technology was swift and, as is typical with disruptive technologies, reactive and somewhat fragmented.[87] Early court decisions regarding elevators grappled with how to fit the new passenger elevator into existing doctrinal frameworks, especially those governing master-servant liability, negligence, and premises liability.[88] James Avery Webb's treatise, *The Law of Passenger and Freight Elevators*, went through two editions between 1896 and 1905,[89] carefully documenting and systematizing the judiciary's efforts to integrate the invention into the existing legal structure.[90] Elevator construction, it was explained, was to be undertaken with adequate care, by using "materials and workmanship . . . such as will stand the work for which it is intended" and "exercis[ing] due diligence in making the elevator reasonably safe."[91] Industry custom regarding elevator construction would be relevant to that determination, but "the usage of others is not the sole criterion," and compliance with industry custom would not support a conclusion as a matter of law that a defendant acted with due diligence.[92]

---

[82] *Id.* at 17 (discussing the Haughwout store installation).

[83] *See* Bernard Andreas, Lifted: A Cultural History of the Elevator 13 (2014); Goodwin, *supra* note 75, at 17 (explaining that even with Otis's safety hoist, the concept was too novel and the machinery too noisy, so many customers "preferred to trust the stairs").

[84] Jacopo Prisco, *A Short History of the Elevator*, CNN (Feb. 9, 2019), https://www.cnn.com/style/article/short-history-of-the-elevator [https://perma.cc/4MU4-CC8E].

[85] Early legal challenges show public apprehension of the devices and efforts to subject them to special legal treatment. *See, e.g.*, Ziemann v. Kieckhefer Elevator Mfg. Co., 63 N.W. 1021, 1023 (Wis. 1895) (rejecting plaintiff's contention that elevators are "imminently dangerous to the lives or persons of others," reasoning that they "are in such universal use that we cannot say that one [is] . . . imminently dangerous to the lives of others," and affirming dismissal of products-liability-type claim for lack of privity between plaintiff and the elevator manufacturer); Donnelly v. Jenkins, 58 How. Pr. 252, 254–55 (N.Y. Com. Pl. 1880) (declining to adopt plaintiff's theory that an elevator in a commercial building "was, in itself, a nuisance," and concluding as a matter of law that plaintiff could not succeed on his negligence theory where no evidence showed how the door of the elevator shaft into which he fell had come to be open).

[86] *See* Goodwin, *supra* note 75, at 27 (describing how "[n]ewspapers . . . entertained their readers with lurid accounts of [elevator] deaths and injuries on the one hand and demands for reform on the other").

[87] *Cf. id.* at 123–24 (discussing inconsistent adoption and enforcement of building-height restrictions in New York, Chicago, and elsewhere, driven by concerns that the tall new buildings that the elevator enabled, even at a mere twelve stories, "blacked out light and flung their denizens all at once into the streets at night").

[88] *See* James Avery Webb, The Law of Passenger and Freight Elevators §§ 35, 59–62, 95, 104–07, 115, 123–28, 143–47 (2d ed. 1905).

[89] *See id.* at v (discussing the need for a second edition in a short span of years).

To any lawyer or student of the law, these principles are eminently familiar: a party must act with reasonable care to avoid causing physical harm to foreseeable victims of her carelessness.[93] That is true whether it is piles of hay,[94] schoolhouse kicks in the shin,[95] or newfangled elevators at issue. Thus, it was largely the general principles of the common law, as elaborated by the courts, that governed the early use of elevators. And yet the very existence of Webb's specialized legal treatise at such an early stage underscores the public's uneasiness with the new technology and the perceived necessity for specially tailored rules.[96]

As is so often the case, generally applicable principles of law did not assuage public concern, and legislatures soon advanced more particular lawmaking to restrict the new devices. An avalanche of legislation followed. Early laws often focused on highly granular operational details, such as rope tensile strength, brake mechanisms, and the qualifications of elevator operators, reflecting a preoccupation with the visible mechanical aspects of safety.[97] Even elevator operators were deemed at risk. In many states, it was made illegal for someone under the age of eighteen to operate an elevator that traveled faster than 200 feet per minute (just over 2.2 miles per hour).[98] New York state law prohibited women from working as elevator operators before 7:00 AM and required they be provided seats, apparently to protect the delicate sex[99] from the dangers of early-morning labor.[100] And Minnesota adopted a (slightly) more nuanced statute, which required that elevators be operated by licensed attendants, but only in cities with populations over 50,000—a law surely satisfying to the Elevator Operators' union,[101] but which apparently conceded that rural folk were hardy enough to look after themselves.[102]

---

[90] *Id.*

[91] *Id.* § 15 (stating a high standard for construction and maintenance of elevators, restricting employers' ability to shield themselves from liability, and using a "reasonably prudent man" standard in determining an owner's responsibility of employing due diligence).

[92] *Id.*; *cf.* The T.J. Hooper, 60 F.2d 737, 737 (2d Cir. 1932) (Hand, J.) ("Indeed in most cases reasonable prudence is in fact common prudence; but strictly it is never its measure; a whole calling may have unduly lagged in the adoption of new and available devices.").

[93] Restatement (Third) of Torts: Liability for Physical & Emotional Harm § 3 (Am. L. Inst. 2010) (negligence where person's conduct "lacks reasonable care" as to "the foreseeable likelihood that the person's conduct will result in harm"); Richard A. Epstein & Gregory M. Dickinson, The Law of Torts § 5.1 (2d ed. 2025) (negligence liability where defendant's conduct "falls below the level of care required of a reasonable person and thereby creates an unreasonable risk of danger to the plaintiff" and that risk of harm is actualized, causing "death, bodily injury, or property damage").

[94] Vaughan v. Menlove, 132 Eng. Rep. 490, 490–94 (C.P. 1837) (affirming liability for negligently maintaining a hayrick despite repeated warnings of the fire hazard and explaining that negligence is measured by "the caution which would have been observed by a man of ordinary prudence" rather than the defendant's own subjective judgment).

[95] Vosburg v. Putney, 50 N.W. 403, 404 (Wis. 1891) (liability for playful but unwelcome and unwarranted schoolhouse shin kick that would foreseeably cause some harm even if not the full measure that actually resulted).

[96] *See also* Webb, *supra* note 88, § 15 (special statutes requiring "safety clutches" and "automatic doors"), § 33 (fire escapes); *cf.* Frank H. Easterbrook, *Cyberspace and the Law of the Horse*, 1996 U. Chi. Legal F. 207–16 (1996) (supporting law-school curriculum focused on the general principles of the law rather than focused study on the law governing particular technologies).

Initial public apprehension, fueled by a combination of genuine risk and an understandable fear of the unknown, compelled legislative bodies to create specific, detailed laws. The tendency towards overly prescriptive and fragmented legislation, often based on an incomplete or shortsighted understanding of the technology or an overreaction to perceived dangers, resulted in cumbersome, rapidly outdated, and sometimes comically misguided legal requirements. The story of the elevator thus serves as a cautionary tale to those who would contemplate particularized enactments to govern today's rapidly evolving technologies.

### III. AI Technologies and the Law's Static Ideal

None of this means that new technology does not sometimes require legal innovation. Very often it does.[103] Typically, however, there will be no need to change substantive law, merely to clarify, where it is not already obvious, how the new technology fits within the existing legal framework. Think, for example, of the torts of conversion and assault. Whether a tortfeasor takes a plaintiff's property by sword or by 3D-printed ghost gun matters not in the least, for the wrongfulness of assault and conversion lies in putting the plaintiff in fear of her safety and depriving her of her property and liberty, not the employment of any particular tool in doing so.[104]

---

[97] *See* Webb, *supra* note 88, § 33. The primary danger of such laws is not that legislators will set unwise safety standards (although they might do that too), but that the lawmaking process is prone to numerous problems discussed *infra* Part IV, such as regulatory capture and ossification that locks in soon-to-be-outdated technologies.

[98] *See id.* §§ 273–74, 277.

[99] Apparently, women of the day disagreed. *Compare Signs Work-Limit Bill. Affects Woman Elevator Operators—Measures Approved by Smith*, N.Y. Times, May 13, 1919, at 16 (explaining that the "Governor approved the measure because he believed that the conservation of the health of women and minors who have to work is an important duty of the State"), *with Women's Work Limited by Law*, N.Y. Times, Jan. 18, 1920, at X4 (reporting complaints by women that in actuality the law was introduced to disqualify them from work in favor of men).

[100] Act of Sep. 1, 1919, ch. 544, §§ 176–77, 1919 N.Y. Laws 1494 (restricting the hours of women elevator operators and requiring seats to be provided for women to protect their health).

[101] *See New Union Has Many Worthy Purposes*, St. Paul Globe, May 10, 1902, at 4 (describing the growth of the union, which by 1902 represented "90 per cent of the elevator men of the city," which the operators said "benefitted themselves to a very great extent," although they also believed that "the public generally will be benefited," and explaining that the union's "next move will be to prohibit the operation of elevators by those without a license").

[102] Act of Apr. 10, 1901, ch. 195, § 1, 1901 Minn. Laws 271 (requiring licenses for elevator operators in cities with populations of over 50,000).

[103] *See, e.g., infra* Part III (discussing large language models, autonomous vehicles, and online fraud). Another leading example is the tort of conversion, which has gradually extended beyond its initial scope to include even digital property. *See generally* Thyroff v. Nationwide

### A. Large Language Models and Defamation

To take on what may seem like a trickier case, consider the emergence of large language models that can engage in human conversation, produce written content indistinguishable from human-authored material,[105] and even pass the bar exam.[106] They also sometimes err and make false and disparaging statements that can cause significant reputational harm. Two leading examples are the mistaken assertion by OpenAI's ChatGPT that Brian Hood, a mayor in Australia, had been imprisoned for bribery,[107] and by Meta's Llama that political activist Robby Starbuck participated in the January 6, 2021, riot in Washington, D.C.[108]

It may seem at first that such a technological advance would require a correspondingly significant legal innovation. Who, before now, ever heard of defamation by machine? Yet, as Eugene Volokh has discussed in a pair of

---

Mut. Ins. Co., 864 N.E.2d 1272 (N.Y. 2007) (discussing the history of the tort, the division of authority on the point, and allowing conversion claim for digital property).

[104] *See* Epstein, *supra* note 6, at 253–56 (reasoning that "the importance attributable to changing social conditions as a justification of new legal doctrines is overstated and quite often mischievous" and offering fraud and duress as an example of where the law has and should remain constant); *cf.* ARISTOTLE, NICOMACHEAN ETHICS 1134b25–26 (W.D. Ross trans.), *in* 2 THE COMPLETE WORKS OF ARISTOTLE 1791 (Princeton Univ. Press 2014) (observing that certain fundamental principles of law are "unchangeable and ha[ve] everywhere the same force (as fire burns both here and in Persia)").

[105] *See* Cameron R. Jones & Benjamin K. Bergen, *Large Language Models Pass the Turing Test* (Mar. 31, 2025) (unpublished study), https://arxiv.org/pdf/2503.23674 [https://perma.cc/3BHY-478T] (showing multiple LLMs capable of passing the Turing test more than 50% of the time by appearing to have a human-like persona).

[106] *See* Karen Sloan, *Bar Exam Score Shows AI Can Keep Up with 'Human Lawyers,' Researchers Say*, REUTERS (Mar. 15, 2023), https://www.reuters.com/technology/bar-exam-score-shows-ai-can-keep-up-with-human-lawyers-researchers-say-2023-03-15/ [https://perma.cc/ZV6F-W7S9] (discussing research finding AI model scores high enough score to pass bar exam and place in 90th percentile of test takers).

[107] Tom Gerken, *ChatGPT: Mayor Starts Legal Bid Over False Bribery Claim*, BBC (Apr. 6, 2023), https://www.bbc.com/news/technology-65202597 [https://perma.cc/2QPX-R2DD].

[108] Sarah Nassauer & Jacob Gershman, *Activist Robby Starbuck Sues Meta Over AI Answers About Him*, WALL ST. J. (Apr. 29, 2025), https://www.wsj.com/tech/ai/activist-robby-starbuck-sues-meta-over-ai-answers-about-him [https://perma.cc/EY4D-QGN6]; Complaint at 1–5, Robert Starbuck v. Meta Platforms, Inc., No. N25C-04-283-SKR-CCLD (Del. Super. Ct. Apr. 29, 2025) (alleging that Meta's "Llama" large language model falsely asserted Starbuck participated in the January 6 Capitol riot and was arrested, detailing his denials, Meta's alleged refusal to correct the record, and the reputational and personal harms caused by the AI's repeated publication of the accusations).

recent articles, even here the long-standing law of defamation provides the necessary tools for analysis.[109] No new legal rights or categories (and no legislation or regulation) are required.[110] That is so because common-law liability for defamation has never required that the defendant have penned or typewritten the defamatory words herself or even that she have spoken or written any words at all.[111] Rather, liability arises when a defendant publishes or continues to publish a defamatory statement about the plaintiff while possessing the required mental state[112] (typically at least negligence under modern law).[113] It does not matter whether the mistake is the result of a human mental slip or the careless operation or design of a machine subject to her control.[114] Either will suffice because it is culpability and publication, not human authorship, that ground the tort.[115]

Applying these principles to LLMs in particular, defamation law will impose liability on an LLM creator if publication of the defamatory content was the result of a defective product design.[116] Similarly, liability will arise where the LLM's design is not defective (that is, no superior product design was available at the time of the product's design), but where the LLM creator is made aware of the defamatory content but nonetheless continues to publish it despite an opportunity to correct the error.[117] Existing defamation principles already govern even the novel case of LLM defamation because the law does not depend on the origin of the statement—whether composed by a person, copied from a wire service,[118] or generated by code—but on the human actor's design error or knowledge and continued publication. The facts may be unfamiliar, but the legal inquiry is not. The common law of defamation is already structured to do the work.

---

[109] *See* Eugene Volokh, *The Duty Not to Continue Distributing Your Own Libels*, 97 Notre Dame L. Rev. 315, 325–32 (2021) [hereinafter, Volokh, *Distributing Your Own Libels*]; Eugene Volokh, *Large Libel Models? Liability for AI Output*, 3 J. Free Speech L. 489, 508–09 (2023) [hereinafter, Volokh, *Large Libel Models?*].

[110] *See* Volokh, *Distributing Your Own Libels*, *supra* note 109, at 325–32 (applying long-standing defamation doctrine of publication by ratification to failures to cease making available one's online postings once learned to be defamatory); Volokh, *Large Libel Models?*, *supra* note 109, at 508–09 (reasoning that output of large language model can support defamation claim despite lack of human speaker by analogy to human error in operation or design of machines, on which defamation claims have historically been predicated despite the lack of any human intent to author the statement communicated).

[111] *See* Restatement (Third) of Torts: Defamation §§ 1–2 (Am. L. Inst., Proposed Draft No. 4, 2024) (requiring only intentional or negligent communication of a defamatory statement, "with fault," not authorship).

[112] Historically the rule was strict liability. *See* Restatement (First) of Torts § 580 (Am. L. Inst. 1938). The modern trend is to require some level of culpability, ranging from simple negligence to recklessness, depending on the subject matter of the statement and its target. For detailed discussion, *see* Restatement (Third) of Torts: Defamation § 1 & cmt. i (Am. L. Inst., Proposed Draft No. 4, 2024); Robert D. Sack, Sack on Defamation §§ 1.2–5, 2.1 (Practising L. Inst. 2025); and Epstein & Dickinson, *supra* note 93, § 18.7.

[113] Restatement (Third) of Torts: Defamation § 1(c) (Am. L. Inst., Proposed Draft No. 4, 2024) (requiring "fault (amounting at least to negligence) with respect to the statement's falsity").

[114] *See, e.g.*, S. Bell Tel. & Tel. Co. v. Coastal Transmission Serv., Inc., 307 S.E.2d 83 (Ga. Ct. App. 1983) (libel liability for errant printing in phonebook of auto repair shop's slogan as "Get it in rear" instead of "Get it in gear"); Volokh, *Large Libel Models?*, *supra* note 109, at 508–10.

[115] *See* Restatement (Third) of Torts: Defamation §§ 1–2 (Am. L. Inst., Proposed Draft No. 4, 2024).

[116] *See* Volokh, *Large Libel Models?*, *supra* note 109, at 491–94, 508–09 (reasoning that common-law defamation principles extend to false statements generated by LLMs, since liability turns on culpability and publication rather than human authorship, and noting that "errors in what a company communicates can be defamatory regardless of whether the errors stem from direct human error in composing text or from human error in creating the technology that produces the text").

### B. Liability for AI Systems

The same is true in tort law more broadly. Questions about accountability for AI systems—whether in cars, hospitals, or homes—are not categorically different from questions courts have answered for centuries. They require adaptation, not fresh lawmaking. Consider harms caused by AI-driven products, such as self-driving cars, virtual assistants, automated medical equipment, for which negligence and products liability law, as discussed below, already offer the basic legal tools.

To take one prominent example, when an autonomous vehicle causes an accident, courts need not depart from long-standing doctrine to assign responsibility. If the incident can be traced to flawed design, programming, or inadequate instructions, standard products liability rules suffice.[119] Similarly, accidents traceable to error by human drivers of semiautonomous vehicles are well-governed under standard negligence law.[120] In both contexts, the underlying principles remain consistent with existing law: manufacturers bear responsibility for defects,[121] users for harms caused by negligent use,[122] and courts and juries for evaluating reasonableness under the circumstances.[123]

---

[117] *See* Volokh, *Distributing Your Own Libels*, *supra* note 109, at 319–27 (concluding that under existing common-law principles, a publisher who learns that its own statement is false and defamatory has a duty to remove or correct it, likening continued online publication to keeping graffiti displayed on one's property after learning of its presence and defamatory character).

[118] For a discussion of the evolution of the wire service defense against strict republication liability and its modern analogue, Section 230 intermediary immunity, see generally Gregory M. Dickinson, *Section 230: A Juridical History*, 28 STAN. TECH. L. REV. 1, 9–11 (2025); Brent Skorup & Jennifer Huddleston, *The Erosion of Publisher Liability in American Law, Section 230, and the Future of Online Curation*, 72 OKLA. L. REV. 635, 638–39 (2020).

[119] *See* David C. Vladeck, *Machines Without Principals: Liability Rules and Artificial Intelligence*, 89 WASH. L. REV. 117, 129–41 (2014) (established products liability principles are well-suited to address harms caused by autonomous machines). *See generally* RESTATEMENT (THIRD) OF TORTS: PRODUCTS LIABILITY § 2 (AM. L. INST. 1998) ("A product is defective when, at the time of sale or distribution, it contains a manufacturing defect, is defective in design, or is defective because of inadequate instructions or warnings.").

[120] *See* RESTATEMENT (THIRD) OF TORTS: LIABILITY FOR PHYSICAL & EMOTIONAL HARM § 3 (AM. L. INST. 2010) (setting forth general duty of reasonable care).

[121] *See* RESTATEMENT (THIRD) OF TORTS: PRODUCTS LIABILITY § 12 (AM. L. INST. 1998).

[122] *See* RESTATEMENT (THIRD) OF TORTS: LIABILITY FOR PHYSICAL & EMOTIONAL HARM §§ 3–4, 6 (AM. L. INST. 2010).

[123] *See id.* § 8.

That framework functions effectively wherever harm can "fairly be attributed to some act or omission by a human that can be said to have 'caused' the accident."[124]

At the same time, AI does introduce new challenges that require careful development of our system's legal principles. For example, some machine behaviors, including those of autonomous vehicles, may defy easy attribution of fault either to operator or manufacturer, particularly where no single design flaw or programming misstep can be pinpointed.[125] Consider a self-driving car that abruptly swerves to avoid a phantom obstacle detected by its sensors, colliding with another vehicle. Was the harm caused by a coding choice, by imperfect training data, or by an unforeseeable interaction of multiple subsystems? Likewise, an autonomous truck might continue accelerating into a traffic jam because its deep-learning navigation system misclassified stopped cars as background scenery—an error traceable only to the inscrutable weightings of a neural network rather than to any discrete programming decision. Other scenarios extend beyond the roadway: an automated rideshare fleet could collectively reroute into a dangerous area due to flawed real-time mapping data, or a robotic delivery vehicle might fail to recognize a pedestrian in low light until too late. Each case involves a black-box chain of perception and action where traditional fault-finding tools falter.

Indeed, courts have already begun to confront these difficulties. In *Huang v. Tesla, Inc.*, litigation followed the death of a driver whose Tesla, operating on autopilot, steered into a highway barrier; while plaintiffs argued the crash reflected a design defect in Tesla's semi-autonomous system, the company maintained that driver misuse and inattentiveness were to blame.[126] Similar issues surfaced in *Nilsson v. General Motors LLC*,[127] where a motorcyclist struck by a self-driving test vehicle alleged negligence, but the precise role of the human safety driver versus that of the autonomous system was hotly contested. Even detailed investigation, such as that which followed Uber's fatal 2018 Arizona crash involving a self-driving test vehicle, may be insufficient to determine whether causation flowed from failures of human oversight, inadequate sensor programming, or systemic flaws in AI systems.[128]

---

[124] Vladeck, *supra* note 119, at 141 (reaching this conclusion after proposing and analyzing application of traditional negligence doctrine to careless human conduct involving semiautonomous systems).

[125] *See id.* at 141–50 (discussing potential approaches for accommodating such cases into existing tort doctrine); *see also* Andrew D. Selbst, *Negligence and AI's Human Users*, 100 B.U. L. Rev. 1315, 1325–30 (2020) (detailing the challenges that AI's inscrutability may pose to fault-based compensation regimes).

[126] *See* Huang v. Tesla, Inc., No. 19CV346663, 2022 Cal. Super. LEXIS 61545 (2022).

[127] *See* Complaint for Damages, Nilsson v. General Motors LLC, No. 18-cv-471 (N.D. Cal. 2018) (No. 4:18-cv-00471-KAW); Answer and Demand for Jury Trial, Nilsson v. General Motors LLC, No. 18-cv-471 (N.D. Cal. 2018) (No. 4:18-cv-00471-JSW).

[128] *See* Nat'l Transp. Safety Bd., Collision Between Vehicle Controlled by Developmental Automated Driving System and Pedestrian v–vi (2019) (outlining the probable causes of the crash as determined by the National Transportation Safety Board).

These disputes underscore how negligence and design-defect frameworks strain when the causal chain runs through opaque algorithmic processes rather than identifiable human error. How can negligence or design-defect products liability law govern where fault cannot be identified and assigned? Yet even in such cases, the law's basic structure is resilient, offering a menu of possible approaches.[129] One option, of course, is to change nothing. Not every person who suffers an automobile accident is entitled to a compensatory legal award, only those who are injured by another's wrongful conduct.[130] If an injured party has not been wronged or cannot prove that she has been wronged, there is no legal basis for recovery from a third party.[131] However, that rule can produce harsh results where we intuit that the defendant must have made some mistake or should bear liability regardless, but AI's complexity makes the factual details of any negligence murky, which precludes the plaintiff from satisfying her burden of proof.

For such cases, the common law offers other paths. Since the mid-twentieth century, for example, tort law has accommodated plaintiffs' difficulty proving product manufacturers' negligence by adopting a strict products liability regime that reduces or eliminates the burden to prove negligence where a plaintiff is injured by a defective product.[132] Over time, "strict" products liability doctrine has drifted toward a negligence standard, at least when applied to design and warning defects, where the plaintiff must show a reasonable alternative design or failure to provide an adequate warning.[133] But a truly strict liability regime (like that applied to manufacturing defects) that imposes liability without proof of fault remains a viable path for AI systems that cause physical injuries to their users or bystanders and has the recommendation of some leading scholars in the field.[134]

---

[129] Indeed, the debate over strict versus negligence-based liability "ha[s] been made in exactly the same fashion both in modern and Roman times," producing an extraordinarily well-developed body of law and legal thought. Epstein, *supra* note 6, at 258–59 nn.15–16 (1980) (collecting sources).

[130] *See* Restatement (Third) of Torts: Liability for Physical & Emotional Harm §§ 5–6 (Am. L. Inst. 2010) (imposing liability for physical harms that were factually caused by an actor's wrongful conduct). Of course, a regime of strict liability for physical harms is possible, perhaps even desirable given reduced litigation costs, but that is not the direction that tort law has taken in most contexts. For discussion of the historical debate and the relative merits of each approach, see generally Epstein & Dickinson, *supra* note 93, §§ 3.1–4.5.

[131] The lack of fault and therefore legal recourse for recovery may seem to present a problem by leaving those injured by AI systems without recovery for their injuries. However, although no-fault legal liability is one way to address such unexpected losses, the more efficient solution is private insurance, which is already well-established in the context of automobile accidents and medical expenses and capable also of protecting individuals against unexpected harms caused by AI systems.

[132] *See* David G. Owen, *Products Liability Law Restated*, 49 S.C. L. Rev. 273 (1998) (classic exposition of the doctrine's history); William L. Prosser, *The Assault Upon the Citadel (Strict Liability to the Consumer)*, 69 Yale L.J. 1099 (1960) (seminal critique of the common-law rules); A. Mitchell Polinsky & Steven Shavell, *The Uneasy Case for Product Liability*, 123 Harv. L. Rev. 1437 (2010) (economic analysis). For a recent, contrarian history that sees products liability as less of a departure from prior law, see generally Alexandra D. Lahav, *A Revisionist History of Products Liability*, 122 Mich. L. Rev. 509 (2023).

[133] For further discussion of this point, *see* Epstein & Dickinson, *supra* note 93, §§ 16.10–16.13.

Another potential legal tool for analyzing liability for AI systems is the doctrine of res ipsa loquitur, which presumes the defendant's carelessness and shifts the burden of proof to the defendant to show that she acted with adequate care if the accident is the sort of thing that ordinarily happens only as a result of negligence by someone in the defendant's position.[135] The doctrine is not only for law-school hypotheticals and pedestrian barrel-hoisting injuries,[136] but serves a crucial burden-shifting role in many modern medical malpractice[137] and products liability[138] cases where direct evidence of negligence or causation is difficult to obtain. The res ipsa loquitur doctrine has proven useful in products liability cases involving vehicles' automated driving features, most prominently in the massive litigation that followed numerous instances of automatic, sudden acceleration of Toyota vehicles in the 2010s.[139]

Finally, it is even possible to carve a middle ground between strict liability and negligence-based fault that does not require inquiry into the sometimes-inscrutable workings of AI systems. Mihailis Diamantis has, for example, proposed a hybrid negligence standard that would gauge the reasonableness of AI performance in comparison to both humans and other algorithms.[140]

---

[134] *See* Matthew Wansley, *The End of Accidents*, 55 U.C. Davis L. Rev. 269, 276 (2021) (suggesting strict liability for all crashes in which an automated vehicle crashes, regardless of operator error); Vladeck, *supra* note 119, at 146 (proposing a strict liability framework for injuries caused by fully autonomous machines in cases where fault cannot be assigned); Selbst, *supra* note 125, at 1324–26 (noting the difficulty that plaintiffs may face in proving reasonable alternative designs and advocating for a regime that includes no such requirement). *But see* Mihailis E. Diamantis, *Vicarious Liability for AI*, 99 Ind. L.J. 317, 331–32 (2023) (noting the limitations of a strict-liability approach for AI systems, which parallel criticisms of strict liability in other contexts).

[135] *See* Restatement (Third) of Torts: Liability for Physical & Emotional Harm § 17 (Am. L. Inst. 2010); Epstein & Dickinson, *supra* note 93, § 7.3.

[136] The seminal case is Byrne v. Boadle, 159 E.R. 299, 300–01 (Ex. 1863) (pedestrian passing below window of flour dealer inexplicably struck by falling barrel).

[137] The classic modern example is Kambat v. St. Francis Hosp., 678 N.E.2d 456 (N.Y. 1997) (physician left a laparotomy pad in a patient's abdominal cavity after concluding a hysterectomy, but the patient lacked direct proof of how it had come to be there). For more recent uses of the doctrine, see, e.g., Barber v. Manatee Mem'l Hosp., Ltd. P'ship, 388 So.3d 279, 287–92 (Fla. Dist. Ct. App. 2024) (sedated, unconscious patient who suffered bilateral hip fractures); Burleson v. Wayne, 495 P.3d 146, 152 (Okla. Civ. App. 2021) (breast implant deflated during rib-graft surgery).

[138] *See, e.g.*, Russell v. Whirlpool Corp., 702 F.3d 450, 458–59 (8th Cir. 2012) (circumstantial evidence was strong enough to support reasonable inference that refrigerator was cause of fire under res ipsa loquitur theory); *cf.* Pavoni v. Chrysler Grp., LLC, 789 F.3d 1095, 1098–99, (9th Cir. 2015) ("[p]roof of malfunction of a part for which auto manufacturer could alone be responsible" may support finding of causation) (citing William L. Prosser, Law of Torts: Products Liability, Proof § 103 (4th ed. 1971)).

[139] *See* In re Toyota Motor Corp. Unintended Acceleration Mktg., Sales Pracs., & Prods. Liab. Litig., 978 F. Supp. 2d 1053, 1080 (C.D. Cal. 2013) (allowing MDL to proceed against Toyota even though plaintiff's expert was "unable to identify with certainty a precise software bug (or other specific cause) that can open the [vehicle's] throttle").

[140] *See* Mihailis E. Diamantis, *Reasonable AI: A Negligence Standard*, 78 Vand. L. Rev. 573, 602–03 (2025) (proposing hybrid negligence standard).

Under his approach, "an AI engaged in some task is reasonable if it causes less harm than the weighted average of harm that all actors—both AI and human—cause while engaged in that task."[141] This formulation would preserve negligence law's existing fault-based structure while updating the comparator class to reflect AI's nonhuman character.

The point here is not to press for any particular solution. Many could work, and there may be a role for several schemes depending on context. The broader point is that despite the novelty of AI systems, the problems of protecting against unavoidable, but unpredictable losses, and of assigning legal liability under uncertain and difficult facts are not new ones. They do not require rethinking from the ground up. Existing law offers a bevy of options, all long studied, with well-understood tradeoffs, and in some instances centuries of judicial experience in application.[142]

### C.  *Dark Patterns, Deep Fakes, and Consumer Protection*

A third frontier of digital technology that might initially seem to require a radical rethinking of the law is digital commerce, which today is saturated with what are sometimes called "dark patterns"—user interface designs that pressure or manipulate users into making choices they would not otherwise make.[143] These designs may obscure relevant information, exploit behavioral biases, or impede users' ability to decline options disfavored by the app maker or website.[144] However, although the phrase is relatively new, the practice is not.[145] From a legal standpoint, these tactics fall squarely within long-standing consumer protection law, including common-law fraud, the FTC Act,[146] and state consumer protection statutes.[147]

---

[141] *Id.* at 603.

[142] *See generally* William M. Landes & Richard Posner, The Economic Structure of Tort Law 2–3 (1987) (describing the centuries-long development of tort liability rules within the common law); Steven Shavell, Foundations of Economic Analysis of Law 1 (2004) (explaining that various tort liability doctrines have long been studied in terms of their effects and tradeoffs).

[143] For detailed discussion of dark patterns, the technologies that power them, and the corresponding rise in online fraud, see Gregory M. Dickinson, *Privately Policing Dark Patterns*, 57 Ga. L. Rev. 1633, 1649–61 (2023). *See generally* Gregory M. Dickinson, *The Patterns of Digital Deception*, 65 B.C. L. Rev. 2457 (2024).

[144] *See* Justin (Gus) Hurwitz, *Designing a Pattern, Darkly*, 22 N.C. J.L. & Tech. 57, 71–72 (2020) (explaining how dark patterns can obscure relevant information and make disfavored choices difficult for users).

[145] *See* Dickinson, *The Patterns of Digital Deception*, *supra* note 143, at 2457, 2464 (observing that "[i]f there are any things new under the sun, human trickery is not among them" and describing how digital deception is only the latest manifestation of commercial deception).

[146] Federal Trade Commission Act, ch. 311, 38 Stat. 717 (1914) (codified as amended at 15 U.S.C. §§ 41–58) (declaring unlawful "unfair or deceptive acts").

[147] For a discussion of these statutes' diverse origins and complicated history along with empirical analysis of their recent applications, see generally James Cooper & Joanna Shepherd, *State Unfair and Deceptive Trade Practices Laws: An Economic and Empirical Analysis*, 81 Antitrust L.J. 947 (2017).

Dark patterns operate by shaping the "choice architecture" presented to users—how options are framed, which defaults are preselected, and which clicks or scrolls are required to make choices.[148] Their methods vary,[149] but their objective is consistent: to shift user behavior in ways that benefit the app maker or website, sometimes at the user's expense.[150] For example, a website might present an interface that requires users to affirmatively decline an option by clicking "No, thanks. I don't like saving money" while highlighting a preferred option in bright green.[151] These techniques range from quite mild to aggressive or downright deceptive.

On the mild end of the spectrum are, for example, forms with the email opt-in box preselected as the default choice (as if that would be most users' preference) and warnings during checkout that a purchase is "unprotected" alongside an offer to buy insurance, sometimes for products the full cost of which barely exceeds the cost of the insurance.[152] These tactics are annoying, but safely navigable, and can be policed by consumers choosing to take their business elsewhere if the annoyance outweighs the benefit of the product or bargain.

On the other end of the spectrum are tactics that achieve their results by confusing, misleading, or outright lying. Such tactics include fake countdown timers ("Only 10 minutes left!") when in reality there is no time limit to a sale; fabricated scarcity alerts ("5 left in stock"); hidden, prechecked boxes for recurring subscriptions; and intentionally confusing double negatives ("Do you not wish to opt out?").[153] Like traditional, in-person sales strategies, these techniques often leverage information asymmetries, user search costs, and deceptive phrasing to influence user behavior.[154] The result is distorted

---

[148] *See* RICHARD H. THALER & CASS R. SUNSTEIN, NUDGE: THE FINAL EDITION 3–5 (rev. ed. 2021) (introducing the concept of "choice architecture" and arguing that the way options are presented can enable policymakers and designers to steer choices).

[149] For the leading taxonomy of dark patterns and the methods they employ, see generally Colin M. Gray, Yubo Kou, Bryan Battles, Joseph Hoggatt & Austin L. Toombs, *The Dark (Patterns) Side of UX Design*, *in* 2018 PROC. CHI. CONF. HUM. FACTORS COMPUT. SYST., Paper 534. For a brief history of dark patterns and a discussion of how app developers can avoid inadvertently introducing them into their products, see generally Gregory M. Dickinson, *Dark Patterns and Consumer Protection Law for App Makers*, 5 PROC. CYBER AWARENESS & RSCH. SYMP. (CARS) (forthcoming 2025).

[150] *See* Hurwitz, *supra* note 144, at 67 (defining dark patterns as designs that steer user behavior toward actions that benefit the designer of the user interface).

[151] Dickinson, *Privately Policing Dark Patterns*, *supra* note 143, at 1637–38 (using this example to illustrate how dark patterns emphasize the designer's preferred choice while making the alternative burdensome or embarrassing).

[152] *See* Laura Daily, *Extended Warranties Benefit Everyone Except the Consumer*, WASH. POST (Nov. 30, 2023), https://www.washingtonpost.com/home/2023/11/28/extended-warranty-waste-of-money/ [https://perma.cc/FKM3-4RM4] (discussing retailers' practice of offering warranties on virtually every product they sell, and concluding "extended warranties are a cost not worth the reward").

[153] *See* Dickinson, *The Patterns of Digital Deception*, *supra* note 143, at 2474, 2485 (discussing the varieties of dark patterns in user interfaces).

[154] *See generally* Michael R. Darby & Edi Karni, *Free Competition and the Optimal Amount of Fraud*, 16 J.L. & ECON. 67, 67–83 (1973) (developing an economic theory of fraud in markets

and inefficient market outcomes: individuals subscribe to services they did not want, pay more than they intended, or consent to data sharing they did not realize was a requirement of the service.¹⁵⁵

None of these tactics, however, is beyond the reach of the law. To the contrary, the common law of fraud already prohibits commercial misrepresentations made with the intent to defraud.¹⁵⁶ The form of the deception—verbal, written, or visual—is immaterial. Modern state and federal consumer protection statutes reach the same result, even eliminating the requirement that the plaintiff prove fraudulent intent. Section 5 of the FTC Act and analogous state consumer protection laws¹⁵⁷ declare unlawful all "unfair or deceptive acts or practices" in or affecting commerce, which the FTC has long interpreted to prohibit any representation "likely to mislead the consumer acting reasonably in the circumstances."¹⁵⁸ Misleading user interfaces fall comfortably within this framework and, indeed, have been a focus¹⁵⁹ of recent FTC actions regarding online sellers.¹⁶⁰

In short, although online fraud has proliferated in recent years, powered by advances in targeted advertising and a general shift to online shopping, lawmaking in this area is a mistake,¹⁶¹ for consumer law is well situated to address the problem. The law's response to dark patterns, like their low-tech forebears, depends on the substance of the defendant's conduct, not the method of deceptive communication. Consumer protection law is already

---

for "credence goods," where consumers cannot easily verify quality even after purchase and explaining how information asymmetries and search costs create opportunities for sellers to mislead buyers in ways that competitive pressures alone may not eliminate).

¹⁵⁵ Dickinson, *supra* note 143, at 1647–49 (explaining how deceptive choice architecture leads consumers to pay more or consent under false impressions, distorting market efficiency).

¹⁵⁶ *See* RESTATEMENT (THIRD) OF TORTS: LIABILITY FOR ECONOMIC HARM § 9 (AM. L. INST. 2020) (providing that "[o]ne who fraudulently makes a material misrepresentation of fact, opinion, intention, or law, for the purpose of inducing another to act or refrain from acting, is subject to liability for economic loss caused by the other's justifiable reliance on the misrepresentation"); EPSTEIN & DICKINSON, *supra* note 93, §§ 20.1–20.13 (discussing the elements of common-law fraud and explaining how the doctrine protects the integrity of market transactions).

¹⁵⁷ *See generally* Henry N. Butler & Joshua D. Wright, *Are State Consumer Protection Acts Really Little-FTC Acts?*, 63 FLA. L. REV. 163 (2011) (tracing the origins of state consumer protection acts as "Little-FTC Acts" designed to supplement governmental enforcement).

¹⁵⁸ Letter from James C. Miller III, Chairman, Fed. Trade Comm'n, to Representative John D. Dingell, Chairman, Comm. on Energy and Com., *FTC Policy Statement on Deception* (Oct. 14, 1983), https://www.ftc.gov/system/files/documents/public_statements/410531/831014deceptionstmt.pdf [https://perma.cc/7YFJ-2LLF].

¹⁵⁹ *See generally* FED. TRADE COMM'N, BRINGING DARK PATTERNS TO LIGHT (2022), https://www.ftc.gov/system/files/ftc_gov/pdf/P214800%20Dark%20Patterns%20Report%209.14.2022%20-%20FINAL.pdf [https://perma.cc/RNB8-WPUZ].

¹⁶⁰ *See* Dickinson, *Privately Policing Dark Patterns*, *supra* note 143, at 1646–53 (describing how the FTC has applied its longstanding authority over "unfair or deceptive acts or practices" to combat harmful interface designs); Hurwitz, *supra* note 144, at 95–100 (noting the FTC's authority over unfair and deceptive acts already covers most harmful dark patterns).

¹⁶¹ For detailed analysis of the shortcomings of actual and proposed laws regarding dark patterns, see Dickinson, *Privately Policing Dark Patterns*, *supra* note 143, at 1649–61.

equipped to do the work. Dark patterns are simply digital tools for the sorts of deception that have never been lawful.[162]

\* \* \*

This Part has shown how even in some of the fastest moving and innovative fields, existing law, while not yet settled in its precise application to those technologies, is well situated to accommodate the new factual scenarios these technologies bring, without fundamental change to substantive law. Far from limiting technological advancement, the law's capacity to absorb new facts without rethinking and rebuilding its foundation is a virtue, for the law thereby continues to guarantee existing rights and freedoms while providing the legal predictability required to encourage investment and innovation.[163] Social change and "variations in individual tastes and conduct can all take place within a persistent set of legal rules."[164]

### IV. Dangers of Technology-Specific Lawmaking

Even if, after reading Parts II and III, you are convinced that the law can evolve slowly and naturally, you might wonder whether we could speed up the process with small doses of technology-specific legislation. Part IV explains why technology-specific lawmaking is often a mistake.

#### A. *Introduction: Particularization and Reconsideration of Generally Applicable Law*

When technological change disrupts the familiar, lawmakers often feel compelled to respond. The elevator, the printing press, and now generative AI have all prompted calls for new legal controls. Yet, new technologies are not born into a legal vacuum; each is born into a full-fledged legal system already in operation. There are, for example, general rules of tort, contract, property, fraud, consumer protection, civil procedure, and due process—time-tested principles widely applied across all manner of contexts and waiting in the wings to govern whatever new technology arises. And so, when legislators act, they do not simply fill a void; they supplant or supplement what already exists. That is, they enact particularized law.

At first blush, particularization may seem prudent. Might statutes tailored to novel technologies govern them more effectively than general-purpose common law? Typically, the answer is no.[165] Particularized law often adds complexity without adding clarity. It raises compliance costs by layering new obligations atop existing ones.[166] It generates inconsistencies by regulating one technological path while leaving nearly identical conduct, accomplished some other way, untouched.[167] Worse, it opens the door to reconsideration of foundational legal principles, often at the behest of whatever special-interest group currently holds political sway.[168] Particularized statutes do not merely coexist with the common law; they place it on the table for negotiation, replacing settled norms with ad hoc compromises that, over time, erode the coherence and generality that make law predictable and just.[169]

---

[162] *Cf.* Hurwitz, *supra* note 144, at 95 (explaining that existing consumer protection law already covers harmful design practices, making new legislative initiatives unnecessary).

[163] For an example of how the natural, coherent extension of the law can go awry when courts or legislatures depart from settled doctrine, see generally Dickinson, *supra* note 118 (discussing the consequences in the context of Section 230 internet immunity of abandoning general-purpose law in favor of a technology-specific framework).

[164] Epstein, *supra* note 6, at 255.

This phenomenon is not hypothetical. As discussed previously,[170] history is replete with examples of lawmakers enacting bespoke rules in the wake of technological panic. From printing presses to comic books to deep fakes, the pattern is consistent. A new tool is feared, then targeted with legislation, often in ways that reflect political alignments more than policy wisdom. These efforts rarely age well. Technologies evolve, use cases proliferate, and the bespoke statute, once cutting-edge, becomes a relic, poorly matched to modern practice but difficult to repeal.[171] The resulting ossification of the legal system, which is stuck trying to regulate a world that no longer exists, reflects what Cass Sunstein calls a structural bias in lawmaking toward overreaction: "[E]lected officers ordinarily face strong incentives to respond to excessive

---

[165] *See supra* Part III.

[166] *See generally* Lachlan Robb, Trent Candy & Felicity Deane, *Regulatory Overlap: A Systematic Quantitative Literature Review*, 17 Reg. & Governance 1131 (2023) (explaining how regulatory overlap can obscure policy objectives and increase compliance costs); *see also* Matthew C. Turk, *Overlapping Legal Rules in Financial Regulation and the Administrative State*, 54 Ga. L. Rev. 791, 856 (2020) (making this point in the context of financial regulation and agency cost-benefit analysis, explaining that "failure to factor in [regulatory] overlap leads the standard CBA procedures astray because it means that they overestimate the benefits of regulatory substitutes [which crowd each other out] and underestimate the benefits of regulatory complements [which amplify one another]").

[167] One contemporary example is Section 230 of the Communications Decency Act of 1996, Pub. L. 104-104, 110 Stat. 56, which provides special protections against vicarious liability only to online entities. *See generally* Gregory M. Dickinson, *Rebooting Internet Immunity*, 89 Geo. Wash. L. Rev. 347 (2021) (noting how, under Section 230, online and offline entities are governed by different rules that immunize only online entities from lawsuits related to content authored by their users or customers).

[168] *See infra* Part IV.C. *See generally* James M. Buchanan & Gordon Tullock, The Calculus of Consent 22–38 (1962) (explaining that public choice theory predicts political actors will favor special interests, allowing group pressures to reshape legal outcomes).

[169] *See* Epstein, *supra* note 7, at 1–49, 151–54 (1995) (making this point generally and using the complexity of modern labor and employment law as an example).

[170] *See supra* Parts II, III.

[171] *See infra* Part IV.E. On the numerous obstacles to lawmaking, including repeal, see William N. Eskridge Jr., *Vetogates and American Public Law*, 31 J.L. Econ. & Org. 756, 757–59 (2015); Michael J. Teter, *Gridlock, Legislative Supremacy, and the Problem of Arbitrary Inaction*, 88 Notre Dame L. Rev. 2217, 2230–31 (2013); Jody Freeman & Matthew C. Stephenson, *The Untapped Potential of the Congressional Review Act*, 59 Harv. J. on Legis. 279, 280–81 (2022).

fear, perhaps by enacting legislation that cannot be justified by any kind of rational accounting."[172]

Even worse, particularized lawmaking poses deeper, institutional risks. The political process that produces new statutes and regulations is not technocratic or dispassionate. It is shaped by interest-group lobbying, skewed incentives, and limited information. Lawmakers act under pressure, often with imperfect understanding, and the laws they produce tend to entrench incumbent interests, privilege visible harms over invisible costs, and congeal into legal regimes that frustrate rather than facilitate innovation.[173] Each new legislative enactment is a wager that our current understanding is sufficient to regulate the future. It rarely is.

The law's strength lies in its generality. Common-law principles endure not because they are forever static, but because they are flexible.[174] They apply across contexts, technologies, and generations, adapting to new facts without requiring wholesale reinvention. Part III makes the case that this generality is a feature, not a flaw—and that technology-specific lawmaking, however well-intentioned, often trades that virtue for a mirage of relevance that quickly fades. In a world of rapid technological change, humility in lawmaking is not weakness, but wisdom.

### B. AI's Knowledge-Problem Paradox

It is tempting, when faced with technological disruption, to reach for the tools of legislation. A problem appears novel, the facts uncertain, and our instincts urge us toward control. The assumption is that new laws will preserve order. But this impulse, however natural, overlooks a profound limitation in human affairs: we do not, and cannot, know enough to legislate wisely, especially in fast-changing environments. This is the essence of the so-called "knowledge problem," first articulated by Nobel-laureate economist Friedrich Hayek in 1937.[175] It is a challenge not of technology, but of epistemology, of the nature of knowledge itself, and nowhere is that challenge more salient today than in legislative efforts to regulate AI.

---

[172] Cass R. Sunstein, *Probability Neglect: Emotions, Worst Cases, and the Law*, 112 YALE L.J. 61, 102–05 (2002) (reasoning that the government should not thoughtlessly regulate in response to highly publicized risks that have low probabilities of occurrence).

[173] *See* Stigler, *supra* note 11, at 3–14, discussed *infra* Part IV.C (developing his public choice theory in which regulation is captured by industries and shaped to serve their own interests, and explaining how political processes tend to privilege concentrated, well-organized groups over diffuse public interests).

[174] *See supra* Part III; Epstein, *supra* note 6, at 253–55 (arguing that the enduring strength of the common law lies in the constancy of its core principles and that legal stability allows private parties to adapt to changing social and technological conditions without the uncertainty that would result from frequent legal change).

[175] *See generally* F.A. von Hayek, *Economics and Knowledge*, 4 ECONOMICA 33 (1937). Hayek expanded on his views in *The Use of Knowledge in Society*, *supra* note 14, now considered the definitive articulation of the concept.

Hayek's knowledge problem begins with a simple insight: "[T]he knowledge of the circumstances of which we must make use never exists in concentrated or integrated form, but solely as the dispersed bits of incomplete and frequently contradictory knowledge which all the separate individuals possess."[176] More specific to the lawmaking context, to design a general welfare-maximizing statute or regulation, a lawmaking body would require access both to scientific information and also to information regarding individual circumstances and preferences. For example, to regulate autonomous vehicles, legislators would need scientific information about the current capabilities and limitations of visual, radar, and lidar sensors; real-time decision-making algorithms; and machine-learning systems.[177] That information would be hard enough to come by on its own, and acquiring and understanding it would risk regulatory capture.[178] But even if it could be acquired, lawmakers would also need granular, localized knowledge about how people in different communities interact with traffic norms, how much risk each individual is willing to tolerate, what tradeoffs each prefers between convenience and safety, and what infrastructural idiosyncrasies—such as local road quality or weather patterns—might affect implementation in each case.[179]

Even when the required scientific knowledge could, in theory, be collected and acted upon, the knowledge of circumstances of time and place is unavailable, and unattainable even in principle. This is not merely a claim about information scale, but about kind. Much of the most valuable knowledge in society is tacit, contextual, and unarticulated.[180] It emerges only through individual trial, error, and adaptation. Even the individuals themselves whose interests are concerned do not know it until circumstances force them to decide. What specific combination of cloud computing resources at a given price point will minimize my company's research costs while maintaining performance that I find acceptable under a fluctuating workload? Or, how much fire-insurance coverage is optimal given my specific home's construction, my local fire department's response times, and my personal risk aversion? Having recently purchased a new home, I now know the answer (or at least I have made a decision). But up until the moment of decision, neither I nor anyone else knew or could have known the answer to that question.

---

[176] Hayek, *supra* note 14, at 519.

[177] For introductory overviews of the relevant technologies, see generally Ed Garston, *What Are Self-Driving Cars? The Technology Explained.*, Forbes (Jan. 24, 2024), https://www.forbes.com/sites/technology/article/self-driving-cars/ [https://perma.cc/T2WC-BYV7]; Center for Sustainable Systems, University of Michigan, *Autonomous Vehicles Factsheet*, Pub. No. CSS16-18 (Oct. 2024), *available at* https://css.umich.edu/sites/default/files/2024-10/Autonomous%20Vehicles_CSS16-18.pdf [https://perma.cc/7HCT-FVZP]; Lauren Sigfussen, *How Do Autonomous Cars Work?*, Discover (Jun. 30, 2018), https://www.discovermagazine.com/autonomous-cars-1670 [https://perma.cc/7SK4-ZT9X].

[178] *See infra* Part IV.C.

[179] *Cf.* Hayek, *supra* note 14, at 521–22 (describing why decision-making requires highly localized, tacit knowledge of conditions such as place, circumstances, and tradeoffs, which cannot be centralized).

[180] *See id.* at 521–27 (explaining that the problem is not merely one of gathering sufficient data, but of using dispersed, tacit, and context-specific knowledge of "the particular circumstances of time and place," which cannot be fully centralized and is instead acted upon via the price system); s*ee also* Ohlhausen, *supra* note 5, at 6–7 (FTC Commissioner discussing the knowledge problem in the context of administrative oversight of online commerce).

The paradox is that even as big data and the AI revolution have made it possible for policy makers to collect more and better information than ever before, central decision makers remain handcuffed by a fundamental inability to obtain the requisite information for optimized lawmaking. No legislature or any other central decision maker—however expert and however advanced its AI—can access, comprehend, and act upon the information that would be needed to guide society's complex interactions. Centrally created rules will necessarily misfire, not out of malice, but out of epistemic impossibility. The relevant information simply cannot be collected and centralized for consideration.

This insight is not merely theoretical. Recent examples illustrate the cost and likelihood of legislating without the requisite knowledge. Public health guidance on alcohol consumption, for instance, has shifted markedly over time—from moderate drinking as heart-healthy, to a focus on cancer risk, to skepticism about benefits of any kind.[181] Even with very comparatively robust econometric data and at least a century of interest in the question, we still do not know whether moderate drinking is good for us, bad for us, or neither.[182]

---

[181] *See* Brianna Abbott & Julie Wernau, *The Dueling Science Behind How Alcohol Affects Your Health*, Wall St. J. (Jan. 25, 2024), https://www.wsj.com/health/alcohol-science-drinking-health-guidelines-aae540e0?gaa_at=eafs&gaa_n=AWEtsqd97xdYd-XvU0vgHOei09yx-AVC1X4ycRG7b4Yfn7ftFTuD8WzSD8XuGhE7LgtM%3D&gaa_ts=68fbf981&gaa_sig=igJKZV8pEgjW6wpG9uGxOGtPfuJJoYd12Yh0M9vCDMC8WIuWZYwXApeOQb0g1TR [https://perma.cc/T3E4-8ZGK] (discussing recent studies on moderate drinking); Roni Caryn Rabin, *Federal Report on Drinking Is Withdrawn*, N.Y. Times (Sep. 5, 2024), https://www.nytimes.com/2025/09/05/health/alcohol-drinking-hhs-report.html [https://perma.cc/6VFK-EPAF]. After decades of study, we have reached the conclusion of antiquity: a bit of vino may be a gift from God, but you can have too much of a good thing. *See* Plato, Laws, *supra* note 66, at 649d–e, p. 1343 (Plato's Athenian noting that he can "point to [no] more suitable pleasure than [drink]—provided some appropriate precautions are taken"). C*ompare* Psalm 104:14–15 ("You make grass grow for the cattle and plants for the people's work to bring forth food from the earth [and] wine to gladden their hearts . . . ."), *and* 1 Timothy 5:23 ("Stop drinking only water, and use a little wine because of your stomach and your frequent illnesses."), *with* Proverbs 23:29–30 ("Who has wounds without cause? Who has redness of eyes? Those who linger long at the wine. . . ."), *and* Luke 21:34 ("Beware that your hearts do not become drowsy from carousing and drunkenness and the anxieties of daily life . . . .").

[182] *Cf., e.g.*, Kevin Shield, Katherine M. Keys, Priscilla Martinez, Adam J. Milam, Timothy S. Naimi & Jurgen Rem, *Draft Report: Scientific Findings of the Alcohol Intake & Health Study for Public Comment,* Substance Abuse & Mental Health Servs. Admin. (Jan. 2025), https://www.samhsa.gov/sites/default/files/2025-draft-public-comment-alcohol-intake-health-study.pdf [https://perma.cc/5RTC-6HP5] (concluding that among both men and women even just one alcoholic beverage per day increases the risk of liver cirrhosis and various cancers); Bruce N. Cologne and Katrina Baum Stone, *Review of Evidence on Alcohol and Health*, Nat'l Acad. Scis., Eng'g & Med. (Sep. 2025), https://nap.nationalacademies.org/read/28582/chapter/1 [https://perma.cc/K62D-5JGC] (concluding moderate alcohol consumption associated with higher risk of breast cancer specifically, but lower risk of cardiovascular mortality and of all-cause mortality); U.S. Dep't Health & Hum. Servs., Alcohol and Cancer Risk: The U.S.

And that is just the scientific question! Truly informed lawmaking would also require information regarding the social benefits to humans of alcohol consumption, differing risk preferences and tolerances between individuals and across circumstances, likely substitute behaviors, and, of course, similarly detailed scientific and particularized information regarding those substitute behaviors.

If we cannot resolve these questions about a behavior as old and studied as drinking, what hope is there of writing wise laws today about generative AI tools? Of course we can guess at what the best answers might be. But there is significant danger in treating temporary guesses as the basis for coercive law. Legislative or regulatory restrictions, once enacted, ossify guesses into rules and punish deviation. Error becomes law.

That risk now looms in the domain of AI. In recent years, many states have enacted laws targeting everything from AI-generated media and deep fakes[183] to dark patterns,[184] algorithmic decision-making,[185] and data-sharing arrangements that make it all run.[186] These laws reflect well-intentioned concerns. But it is far from clear, to put it mildly, that lawmakers possess the

---

Surgeon General's Advisory (2025), https://www.hhs.gov/sites/default/files/oash-alcohol-cancer-risk.pdf [https://perma.cc/6AQC-3LZ8] (concluding that risk of certain cancers starts to increase at one drink or fewer per day, but not discussing potentially offsetting cardiovascular effects or all-cause mortality).

[183] *See, e.g.*, N.H. Rev. Stat. Ann. § 638:26-a (2024) (making it illegal to distribute a deep fake for the purpose of embarrassing, harassing, entrapping, defaming, extorting, or otherwise causing any financial or reputational harm); 2025 N.J. Laws ch. 40 (establishing criminal penalties for production or dissemination of deep fakes); Ariz. Rev. Stat. Ann. § 16-1024 (2024) (preventing a creator from creating and distributing deep fakes of candidates 90 days before an election unless the creation includes a clear and conspicuous disclosure).

[184] *See, e.g.*, California Consumer Privacy Act of 2018, Cal. Civ. Code § 1798.100 (West 2022); Colorado Consumer Privacy Act, Colo. Rev. Stat. § 6-1-1313 (West 2022); Cal. Civ. Code §§ 1798.135(b)(2), (c)(4) (2025) (requiring express re-consent to use of personal information if a consumer has previously opted out, and requiring express consent before an entity may share the personal information of a consumer under sixteen years of age and that the entity "wait for at least 12 months before requesting the consumer's consent again").

[185] *See, e.g.*, Colorado AI Act (S. 24-205), §§ 6-1-1602 to -1604, 6-1-1-1703 to -1704, 2024 Colo. Sess. Laws ch. 205 (enacted May 17, 2024) (to be codified at Colo. Rev. Stat. §§ 6-1-1602 to -1604 and 6-1-1703 to -1704 (2026)) (requiring AI developers to use reasonable care to protect consumers from algorithmic discrimination); 2024 Ill. Laws 103-804 (enacting H.B. 3773, effective Jan. 1, 2026) (to be codified at 775 ILCS 5/2-101(L)(1)–(2) (2026)) (prohibiting discriminatory use of AI in employment decisions); N.Y.C. Local Law No. 144 (2021), N.Y.C. Admin. Code § 20-870 (effective July 5, 2023) (requiring bias audits of AI tools before their use in employment decisions).

[186] *See, e.g.*, Cal. Civ. Code § 1798.140(h) (West 2024) (explaining that consent to data collection cannot be through "[a]cceptance of a general or broad terms of use" or obtained "through use of dark patterns"); Colo. Rev. Stat. §§ 6-1-1303(5)(a), (c) (2024) (same); *cf.* Commission Regulation 2022/2065, of the European Parliament and of the Council of 19 Oct. 2022 on a Single Market for Digital Services and Amending Directive 2000/31/EC (Digital Services Act), 2022 O.J. (L 277) 58 ("[O]nline platforms shall not design, organise or operate their online interfaces in a way that deceives or manipulates the recipients of their service."); *see also* Deceptive Experiences to Online Users Reduction (DETOUR) Act, H.R. 6083, 117th Cong. § 3(a) (2021) (proposed bill that would prohibit user interfaces "with the purpose or substantial effect" of undermining "user autonomy, decision making, or choice to obtain consent or user data").

knowledge to write such rules wisely—or that the benefits of intervention outweigh the costs of suppressing valuable experimentation. Does the value of reducing deep-fake political ads that deceive some voters outweigh the loss of some compelling AI-generated political parody that might have persuaded others, but never made it to the marketplace of ideas for fear of legal liability? No one knows. How about forgone experimentation with intuitive user-interface designs for fear of inadvertently creating a "dark pattern" (whatever that might mean to an aggressive enforcement agency). Or the loss of innovative, not-yet-known apps never created because they cannot be financed without the traditional data-sharing and targeted-advertising revenues that drive the market for online services?[187] Do the gains from dark-pattern and data-privacy legislation outweigh these losses? No one knows that either. But we now have legislation on the books that presumes such knowledge.[188]

Hayek anticipated this danger. The problem with central planning, he argued, is not merely inefficiency, but ignorance.[189] Planners act on the assumption that the relevant knowledge is available to them. But the knowledge required to make wise decisions is not fixed or centralized. It is distributed, situational, and often inarticulable.[190] The proper role of law, then, is not to mandate behaviors based on intuitions and uncertain predictions, but to preserve the conditions under which decentralized discovery can occur.[191] The common law's generality is its virtue. Courts will adapt long-standing legal principles to new technologies without requiring the legislature to anticipate and regulate every innovation.

---

[187] Regarding the negative effects the EU GDPR's data-sharing restrictions on innovation, see generally Guy Aridor, Yeon-Koo Che & Tobias Salz, *The Effect of Privacy Regulation on the Data Industry: Empirical Evidence from the GDPR*, 54 RAND J. Econ. 695 (2023); Garrett A. Johnson, Tesary Lin, James C. Cooper & Liang Zhong, *COPPAcalypse? The Youtube Settlement's Impact on Kids Content*, SSRN (July 10, 2025), https://ssrn.com/abstract=4430334 [https://perma.cc/GQS4-D62K]; Samuel Goldberg, Garrett Johnson & Scott Shriver, *Regulating Privacy Online: An Economic Evaluation of the GDPR*, 16 Am. Econ. J.: Econ. Pol'y 325 (2024); Rebecca Janßen, Reinhold Kesler, Michael E. Kummer & Joel Waldfogel, *GDPR and the Lost Generation of Innovative Apps* (Nat'l Bureau of Econ. Rsch., Working Paper No. 30028, 2022), https://www.nber.org/papers/w30028 [https://perma.cc/8C5T-KVPX]; U.K. Competition & Mkts. Auth., Online Platforms and Digital Advertising: Market Study (2020).

[188] On this ground, Cass Sunstein similarly criticizes the "precautionary principle," which counsels risk aversion rather than risk neutrality in the face of the unknown. *See generally* Sunstein, *supra* note 3.

[189] *See* Hayek, *supra* note 14 (arguing that the problem with centralized rulemaking is not just inefficiency but ignorance of dispersed knowledge).

[190] *See id.*

[191] For a brief articulation of Austrian Law and Economics and Hayek's vision of common-law efficiency as powered by decentralized discovery of efficient solutions, see generally Todd J. Zywicki & Edward P. Stringham, *Austrian Law and Economics and Efficiency in the Common Law*, in Research Handbook on Austrian Law and Economics 192–208 (Todd J. Zywicki & Peter J. Boettke eds., 2017).

### C. Big Tech and Regulatory Capture

In recent years, public anxiety over AI, algorithmic decision-making, and data privacy has fueled calls for legislative and regulatory intervention.[192] Often these demands arise from an understandable concern about safety, fairness, and privacy. But in their haste to protect the public from perceived technological threats, lawmakers may unwittingly expose the public to the subtler and more enduring danger of regulatory capture. The very platforms and app makers whose companies make prompt calls for regulation often play a decisive role in shaping the statutes and rules designed to regulate them, thereby securing legal environments that entrench incumbents, suppress competition, and convert public fears into private moats.

Regulatory capture describes the process by which regulatory agencies or legislative bodies come to be dominated by the industries they are charged with overseeing.[193] In the words of George Stigler, who famously formalized the theory, "as a rule, regulation is acquired by the industry and is designed and operated primarily for its benefit."[194] This does not mean that all regulation is futile or malign. It means rather that regulation will tend over time to reflect the preferences of concentrated industry groups rather than the interests of the general public. That is so because industry groups have both superior resources and greater interest in influencing legislation than the diffuse public.[195]

The tendency toward regulatory capture is amplified in sectors characterized by rapid technological innovation, information asymmetry between industry and regulators, and a high degree of resource concentration, of which modern digital platforms and AI tools are prime examples.[196] Unlike conventional industries, Big Tech operates in spaces where the underlying technologies evolve at breakneck speed and where technical fluency is essential even to frame meaningful questions, let alone draft or implement sound legal rules.

---

[192] *See* Kevin Frazier & Adam Thierer, *1,000 AI Bills: Time for Congress to Get Serious About Preemption*, Lawfare (May 9, 2025), https://www.lawfaremedia.org/article/1-000-ai-bills--time-for-congress-to-get-serious-about-preemption [https://perma.cc/A3HX-V6UF] (observing that 1,000 AI-related bills were introduced in state and federal legislatures in the first four months of 2025).

[193] *See* Ernesto Dal Bó, *Regulatory Capture: A Review*, 22 Oxford Rev. Econ. Pol'y 203, 203 (2006) (explaining that "regulatory capture is the process through which special interests affect state intervention in any of its forms . . . specifically the process through which regulated monopolies end up manipulating the state agencies that are supposed to control them").

[194] Stigler, *supra* note 11, at 3.

[195] *See* Olson, *supra* note 12, at 2 (explaining why large, diffuse publics fail to organize effectively while smaller, concentrated groups can marshal resources to influence legislation); *see also* Sam Peltzman, *Toward a More General Theory of Regulation*, 19 J.L. & Econ. 211, 212–13 (1976) (citing Stigler and arguing that information and organization costs prevent diffuse publics from matching the political pressure of resource-rich industry groups).

[196] *See* Thomas Metcalf, *AI Safety and Regulatory Capture*, AI & Soc'y 3–4 (2025) (explaining why rapid innovation, information asymmetry, and resource concentration make AI and other high-tech industries particularly susceptible to capture).

In such an environment, policymakers face two knowledge deficits. First, they may lack internal expertise.[197] Even well-resourced agencies (or Microsoft, for that matter!)[198] struggle to recruit and retain specialists with up-to-date knowledge of machine-learning systems or emerging targeted-advertising technologies.[199] Second, even if agency staff members possess the relevant technical expertise, they will lack contextual understanding of how it is being deployed in the industry.[200] It is not simply that the government lacks technical knowledge, but that the most relevant information—how users behave, how algorithms are fine-tuned in deployment, and how firms weigh tradeoffs—is local, tacit, and not easily transferred from firm to state.[201]

This knowledge imbalance creates a natural opportunity for the regulated to become the regulators. Policymakers, recognizing their own ignorance, seek expertise from those who have it.[202] But the very actors supplying this expertise—dominant tech companies—are not disinterested. They have powerful incentives to frame problems in ways that support rules they are best positioned to satisfy. Often, they promote regulatory standards that mirror their own internal practices, presenting these as "best practices" or "responsible tech," knowing that competitors with different architectures or fewer resources will struggle to comply.[203]

---

[197] *Cf.* Dal Bó, *supra* note 193, at 214–15 (explaining that agencies often rely on revolving-door hires from industry, creating risks of bias in regulation); Ryan Hagemann, Jennifer Huddleston Skees & Adam Thierer, *Soft Law for Hard Problems: The Governance of Emerging Technologies in an Uncertain Future*, 17 Colo. Tech. L. Rev. 37, 69–70 (2019) (collecting sources and discussing limitations on agencies' acquisition and analysis of information).

[198] *See* Nik Froehlich, *Why It's Difficult to Hire (and Retain) Quality Tech Talent*, Forbes (Jan. 22, 2024), https://www.forbes.com/councils/forbestechcouncil/2024/01/22/why-its-difficult-to-hire-and-retain-quality-tech-talent/ [https://perma.cc/VRH9-CYY6].

[199] *See* U.S. Gov't Accountability Off., GAO-25-108412, Fraud and Improper Payments: Data Quality and a Skilled Workforce Are Essential for Unlocking the Benefits of Artificial Intelligence 9 (Apr. 2025) (noting "a severe shortage of federal staff with AI expertise" and that non-competitive compensation prevents agencies from recruiting and retaining specialists in AI and related fields).

[200] *Cf.* Hayek, *supra* note 14, at 519 (noting that "[t]he peculiar character of the problem of a rational economic order is determined precisely by the fact that the knowledge of the circumstances of which we must make use never exists in concentrated or integrated form, but solely as the dispersed bits of incomplete and frequently contradictory knowledge which all the separate individuals possess").

[201] *Cf.* Ohlhausen, *supra* note 5, at 6 (explaining that regulators must deal with "at least three significant knowledge-gathering challenges. First, a regulator must acquire knowledge about the present state and future trends of the industry being regulated. The more prescriptive the regulation, and the more complex the industry, the more detailed the knowledge the regulator must collect. Second, collecting such information is very time-consuming, if it is even possible, because such knowledge is generally distributed throughout the industry and may even be latent. Third, as a regulated industry continues to evolve, collected knowledge can quickly become stale").

[202] *Cf.* Dal Bó, *supra* note 193, at 214–15, 220–21 (noting that firms offer expertise to regulators, but that expertise is entangled with self-interest, raising risks of bias in regulatory decisions).

[203] *See* Petra A. Nylund & Alexander Brem, *Standardization in Innovation Ecosystems: The Promise and Peril of Dominant Platforms*, 194 Tech. Forecasting & Soc. Change 122714, 5 (2023) (explaining that dominant firms embed their internal processes into standards, present them as best practices, and thereby disadvantage competitors with different architectures

Differing concentrations of economic interest between the public and regulated market participants further encourage regulatory capture. Regulated companies have a very large stake in the content of a new law governing their industry, whereas, although the effects in the aggregate may be large, each voter has comparatively little individual interest.[204] As Sam Peltzman observed, regulation tends to arise when concentrated beneficiaries are able to mobilize for favorable outcomes and when the diffuse public is unlikely to perceive or resist the cost.[205] Big Tech firms are ideally situated to mobilize and deploy resources to acquire beneficial legislation. They command vast lobbying budgets,[206] maintain close relationships with key decisionmakers, and possess both the technical knowledge[207] and legal talent necessary to shape draft language, comment on rules, and litigate the boundaries of interpretation.[208]

A recent episode that illustrates these dynamics involves California's landmark privacy legislation, the California Consumer Privacy Act (CCPA).[209] Initially passed in 2018, the CCPA was widely hailed as a step toward stronger privacy rights in the digital era.[210] But the statute's path to enactment

---

or fewer resources); *cf., e.g.*, OpenAI, Letter to the Office of Science & Technology Policy on AI Action Plan (Mar. 13, 2025), https://cdn.openai.com/global-affairs/ostp-rfi/ec680b75-d539-4653-b297-8bcf6e5f7686/openai-response-ostp-nsf-rfi-notice-request-for-information-on-the-development-of-an-artificial-intelligence-ai-action-plan.pdf [https://perma.cc/3BYB-XB9S] (submitted by Christopher Lehane, Vice President, Global Affairs, outlining OpenAI's regulatory proposals, characterizing a Chinese AI lab DeepSeek as "state-controlled" and recommending the U.S. government ban DeepSeek models); Tripp Mickle, Anna Swanson, Meaghan Tobin & Cade Metz, *U.S. Targets DeepSeek and Its Chips from Nvidia*, N.Y. Times (Apr. 17, 2025), https://www.nytimes.com/2025/04/16/technology/nvidia-deepseek-china-ai-trump.html [https://perma.cc/78K5-3UWJ] (reporting on OpenAI's letter and Trump Administration's response).

[204] *See* Stigler, *supra* note 11, at 10–13 (showing that concentrated industry stakes drive engagement, while dispersed public costs yield apathy).

[205] *See* Peltzman, *supra* note 195, at 213 (noting that "[t]he larger the group that seeks the transfer, the narrower the base of the opposition and the greater the per capita stakes that determine the strength of opposition, so lobbying and campaigning costs will rise faster than group size").

[206] *Cf. 2025 Top-Performing Lobbying Firms: Behind the Data*, Bloomberg (Apr. 2025), https://assets.bbhub.io/bna/sites/20/2025/04/2025-Top-Performing-Lobbying-Firms-Report.pdf [https://perma.cc/L8XV-M372] (noting that "[t]echnology issues were keeping lobbyists busy in 2024 are on track to remain revenue drivers in 2025"); Gary S. Becker, *A Theory of Competition Among Pressure Groups for Political Influence*, 98 Q.J. Econ. 371, 372 (1983) ("Political influence is not simply fixed by the political process, but can be expanded by expenditures of time and money on campaign contributions, political advertising, and in other ways that exert political pressure.").

[207] *Cf.* Dal Bó, *supra* note 193, at 211–12 (surveying the literature explaining "provision of information" by biased advisors as a mechanism for industry to exert influence on regulators).

[208] *Cf.* Stigler, *supra* note 11, at 11–12 (describing how organized interests deploy substantial resources to influence policymakers); Dal Bó, *supra* note 193, at 211–19 (suggesting that resource-rich firms can shape the content and interpretation of rules, in part by deploying superior legislative lobbying capacity).

[209] *See* California Consumer Privacy Act of 2018 ("CCPA"), Cal. Civ. Code §§ 1798.100–1798.199 (2018).

[210] *See, e.g.*, Daisuke Wakabayashi & Issie Lapowsky, *California Passes Sweeping Law to Protect Online Privacy*, N.Y. Times (June 28, 2018), https://www.nytimes.com/2018/06/28/technology/california-online-privacy-law.html [https://perma.cc/8QVJ-C4FU] (discussing

reveals a more complicated story. The legislative process was catalyzed not by deliberate policymaking, but by a threatened ballot initiative funded by a wealthy real estate developer.[211] Lawmakers feared that robust individual privacy protections would be approved by California voters by ballot initiative.[212] In response, legislators hurriedly passed the CCPA to forestall that initiative, relying in part on language and guidance drafted by lobbyists and industry insiders, which softened the law's requirements.[213] The result was a law with vague definitions and exceptions that favored industry participants over voter preferences.[214]

One key provision in early drafts defined "personal information" so broadly that it threatened the business models of many smaller ad tech firms—but left open pathways for dominant platforms to argue that their practices were outside the scope of regulation.[215] Moreover, while the statute imposed new compliance burdens across the board, large firms were better positioned to absorb the costs, which would lead smaller competitors to exit the market or forego certain services altogether.[216] Industry groups lobbied not to kill

---

granting consumers the right to know about company data-sharing); Dan Frank, *New Law May Drive Privacy Strategy Refresh*, WALL ST. J. (Nov. 17, 2020), https://deloitte.wsj.com/cmo/new-law-may-drive-privacy-strategy-refresh-01605643331? [https://perma.cc/AKZ9-9W4C] (denoting expansive provisions to increase transparency); Rachel Lerman, *California Begins Enforcing Digital Privacy Law, Despite Calls for Delay*, WASH. POST (July 1, 2020), https://www.washingtonpost.com/technology/2020/07/01/ccpa-enforcement-california/ [https://perma.cc/3QTS-N9SE] (referencing the law's broad ability to stop selling consumer personal data).

[211] For the history of the CCPA's enactment, see Yunge Li, Note, *The California Consumer Privacy Act of 2018*, 32 LOY. CONSUMER L. REV. 165, 170–72 (2019); Dominique-Chantale Alepin, *Social Media, Right to Privacy and the California Consumer Privacy Act*, 29 COMPETITION J. 96 (2019); *see also* Marc Vartabedian, *California Passes Sweeping Data-Privacy Bill*, WALL ST. J. (June 28, 2018), https://www.wsj.com/articles/california-rushes-to-tighten-data-privacy-restrictions-1530190800? [https://perma.cc/8PVD-PQDC] (noting the ballot initiative led by San Francisco real-estate developer Alastair Mactaggart, which California legislators sought to head off by enacting their own privacy bill).

[212] *See* Alepin, *supra* note 211, at 96 ("The California legislature was under immense pressure to meet this demand and pass legislation—a privacy law passed through the ballot process could prove unworkable both for industry and for consumers. For once a ballot initiative passes and is enacted, it cannot be amended by the state legislature.").

[213] *See* Nicholas Confessore, *The Unlikely Activists Who Took on Silicon Valley—and Won*, N.Y. TIMES (Aug. 14, 2018), https://www.nytimes.com/2018/08/14/magazine/facebook-google-privacy-data.html [https://perma.cc/ABH9-XY6V] (explaining how a lobbyist met with Google and Facebook to propose specific language for the ballot initiative based on failures he denoted).

[214] *See* Salomé Viljoen, *The Promises and Pitfalls of the California Consumer Privacy Act*, DIGIT. LIFE INITIATIVE (Feb. 19, 2021), https://dli.tech.cornell.edu/post/the-promise-and-pitfalls-of-the-california-consumer-privacy-act [https://perma.cc/82RF-5UUL] (discussing some of the weaknesses, from the consumer's perspective, of the CCPA's initial language). The point here is that voter preferences were circumvented, not that they would have made for good policy.

[215] *See* Nicholas F. Palmieri III, *Who Should Regulate Data?: An Analysis of the California Consumer Privacy Act and Its Effects on Nationwide Data Protection Laws*, 11 HASTINGS SCI. & TECH. L.J. 37, 38 (2020) (describing how the definition broadened to include everything that is reasonably capable of being associated with a particular person or household).

[216] *See* CAL. ATT'Y GEN., STANDARDIZED REGULATORY IMPACT ASSESSMENT: CALIFORNIA CONSUMER PRIVACY ACT OF 2018 REGULATIONS (2019) ("Small firms are likely to face disproportionately higher share of compliance costs relative to larger enterprises. Resources explain

the bill, but to shape it—blunting its most aggressive provisions, embedding favorable definitions, and creating a regulatory framework that reinforced their existing advantages.[217]

The same pattern played out in the implementation stage. As enforcement authority transitioned to the newly formed California Privacy Protection Agency, major tech firms actively engaged with regulators, commenting on proposed rules, funding white papers, and hiring former government officials to lead their compliance divisions.[218] At each stage, the process was influenced by a difficult information asymmetry: Regulators relied on the regulated to explain the law's effects, the regulated offered interpretations that advanced their interests, and the resulting explanations hardened into law.[219]

This is not to say that the CCPA was pointless or even that anyone acted in bad faith at any stage of the process. Businesses naturally pressed for lawmakers to craft the legislation in a way that would not undermine their business models. And legislators naturally relied on industry participants' advice to avoid inadvertently crafting a disastrously over-restrictive bill. That is how the legislative process works. And despite business influence, the law did meaningfully expand user rights and even prompt other jurisdictions to consider similar legislation. But the CCPA also exemplifies the capture risks endemic to tech regulation. When lawmakers act without full understanding of the technologies at issue, as they must always do, and when they depend on incumbent firms for technical advice and political support, as, again, they must always do, the resulting laws are likely to bear the marks of their true authors.

The lesson is not that regulation is futile, nor that industry input should be shunned. Rather, the lesson is cautionary. In domains where technical complexity is high and policy knowledge is low, the likelihood of capture rises. Regulators may find themselves implementing the strategic preferences of dominant firms, mistaking them for neutral expertise. And legislation intended to constrain power may instead entrench it. In the end, the choice is not between the status quo and enlightened, welfare-enhancing policy, but between the status quo and whatever Google might wish the law to be.

---

this dichotomy as large technology companies are often several steps ahead of both competitors and regulators.").

[217] *See* Viljoen, *supra* note 214 (explaining how industry lobbying weakened the CCPA by securing exemptions, favorable definitions, and narrowed protections).

[218] *See* Stephen Hobbs, *New California Privacy Agency Faces Pressure From Business, Consumer Groups As It Draws Up Rules*, Sacramento Bee (Feb. 3, 2023), https://www.sacbee.com/news/politics-government/article271944597.html [https://perma.cc/G4WV-3MJE] (explaining how tech companies created advocacy groups to delay enforcement of the CPPA).

[219] *See id.* (describing how legislators relied on input from regulated industry groups, whose interpretations and compromises shaped the final statutory language); Li, *supra* note 211, at 170–72 (also describing how legislators relied on input from regulated industry groups).

### D. The Nirvana Fallacy: Biased Enforcement and Other Principal-Agent Problems

Legislation enacted in response to novel technologies often proceeds as if omniscience exists. Policymakers frame statutes with seeming confidence, treating uncertainty as a problem that can be solved rather than a natural limitation to be respected.[220] But in doing so, they may fall prey to what Harold Demsetz famously called the "Nirvana Fallacy"—the logical flaw of comparing the imperfect present not to a realistically achievable alternative, but to a hypothetical ideal.[221] The fallacy is not merely one of optimism, but of method. It discounts tradeoffs that are implicit within new legislative proposals and hides institutional constraints.[222] Regulatory capture, already discussed,[223] is one such constraint. Others are the risks of biased and self-interested enforcement, discussed below.

#### 1. Bias and Selective Governmental Enforcement

All enforcement regimes face a resource constraint. No agency can investigate or prosecute every technical violation of law.[224] Enforcement is necessarily selective. But selectivity opens the door to bias. That bias may arise from political influence, institutional ideology, or individual discretion.[225] In any case, it raises the risk that rules will be enforced not impartially, but strategically. As Justice Jackson observed decades ago, "[t]he most dangerous power of the prosecutor" is not the power to charge wrongdoing, but that "he will pick people he thinks he should get, rather than pick cases that need to be prosecuted."[226] With the vast, uncountable number of civil and criminal provisions on the books, enforcers can just "pick[] the man, and then search[] the law books . . . to pin some offense on him."[227]

---

[220] *See* Sunstein, *supra* note 3, at 79–85 (explaining that much legislation reflects "probability neglect," citing examples such as Love Canal hazardous waste regulation, Alar pesticide bans, shark-attack legislation, and terrorism laws, all enacted without regard to the actual likelihood of harm); Gary E. Marchant, *Governance of Emerging Technologies as a Wicked Problem*, 73 Vand. L. Rev. 1861, 1863–65 (2020) (explaining that traditional regulatory enactments addressing rapidly advancing technologies "are likely to be outdated before the ink dries," and emphasizing the enormous uncertainty about the trajectories, benefits, and risks of such technologies); *cf.* Barry R. Weingast & Mark J. Moran, *Bureaucratic Discretion or Congressional Control? Regulatory Policymaking by the Federal Trade Commission*, 91 J. Pol. Econ. 765, 791–92 (1983) (finding that the FTC's policies shifted with changes in congressional oversight committees, showing that enforcement reflected political influence rather than neutral administration, another risk that may be unaccounted for).

[221] *See* Harold Demsetz, *Information and Efficiency: Another Viewpoint*, 12 J.L. & Econ. 1, 1 (1969) ("The view that now pervades much public policy economics implicitly presents the relevant choice as between an ideal norm and an existing 'imperfect' institutional arrangement. This *nirvana* approach differs considerably from a *comparative institution* approach in which the relevant choice is between alternative real institutional arrangements.")

[222] *See* Sunstein, *supra* note 3, at 26–34 (arguing that new legislative proposals often create offsetting risks or hidden costs, showing that statutes framed with optimism obscure the real tradeoffs involved).

[223] *See supra*, Part III.C.

[224] One current example is the FTC's limited ability to police dark patterns affecting online commerce, *see* Dickinson, *The Patterns of Digital Deception*, *supra* note 143, at 2483.

[225] *See* William A. Niskanen Jr., Bureaucracy and Representative Government 3–14 (1971) (suggesting that bureaucrats pursue their own preferences within institutional constraints, so enforcement discretion is shaped by ideology and self-interest rather than a neutral public interest).

Public choice theory predicts and explains this result.[228] Agencies, like other political actors, do not pursue pure, abstract goals of justice.[229] They are influenced, sometimes decisively, by interest groups, congressional overseers, and shifting political coalitions.[230] Enforcement decisions are no exception. A regulator's decisions, including whether and whom to prosecute, often reflect personal concerns and powerful economic and political interests, rather than an impartial assessment of public harm.[231]

Discretion in enforcement, in other words, may be a practical necessity, but it is also a deep structural vulnerability. Discretion can be used to protect allies, punish enemies, or advance ideological projects under the guise of neutral law enforcement. For example, Kent Barnett has shown how administrative adjudicators, many of whom are hired, paid, and promoted by the agencies whose actions they are charged with reviewing, can be structurally biased in favor of the government's enforcement priorities.[232] In practice, this means the agency can take enforcement action against disfavored parties while shielding itself or its allies from meaningful review.

---

[226] Glenn Harlan Reynolds, *Ham Sandwich Nation: Due Process When Everything Is a Crime*, 113 COLUM. L. REV. SIDEBAR 102, 103 (2013) (quoting Justice Jackson).

[227] *Id.* at 102; *see* NEIL GORSUCH & JANIE NITZE, OVER RULED: THE HUMAN TOLL OF TOO MUCH LAW 11 (2024) (lamenting the situation, noting that "[t]oo much law amounts to no law at all, for when legal doctrine makes everyone an offender, the relevant offenses have no meaning independent of law enforcers' will") (quoting WILLIAM J. STUNTZ, THE COLLAPSE OF AMERICAN CRIMINAL JUSTICE 3 (2011)); *see e.g.*, United States v. Foreste, 780 F.3d 518, 525–26, 525 n.5 (2d Cir. 2015) (Wesley, J.) (observing that "[g]iven the heavy regulation of automobile travel, it is unlikely that [a police] officer would have much trouble spotting a violation if he set his mind to it" and noting that, in that case, an officer testified that he had succeeded after setting out to intercept a particular vehicle by "find[ing] a motor vehicle violation if at all possible").

[228] For an accessible overview of public choice theory as applied to administrative decision-making, see MAXWELL L. STEARNS, THOMAS J. MICELI & TODD ZYWICKI., LAW AND ECONOMICS: PRIVATE AND PUBLIC 781–806 (2018).

[229] *See* Weingast & Moran, *supra* note 220, at 774–75, 791–93 (emphasizing agency oversight by Congress and hence bureaucrats' concern to further lawmakers' agendas, especially those on powerful committees); NISKANEN, *supra* note 225, at 38 (emphasizing "salary, perquisites of the office, public reputation, power, patronage, output of the bureau, ease of making changes, and ease of managing the bureau" as focuses of bureaucrats' efforts). *See generally* JAMES Q. WILSON, BUREAUCRACY: WHAT GOVERNMENT AGENCIES DO AND WHY THEY DO IT 179–95 (1989) (emphasizing bureaucrats' interest in avoiding risk and conflict with other agencies, which might reduce their autonomy).

[230] *See* Stigler, *supra* note 11, at 3–9 (arguing that regulation is typically acquired and shaped by industry groups for their own benefit, showing the influence of organized interests on lawmaking); Weingast & Moran, *supra* note 220, at 765–93 (finding that FTC policy choices shifted with changes in congressional oversight committees).

[231] *See* M. P. Baumgartner, *The Myth of Discretion*, *in* THE USES OF DISCRETION 157 (Keith Hawkins ed., 1992) ("Left to their own devices, agents of the law routinely favor some sorts of people over others. Discretion, in practice, amounts to what is commonly known as discrimination."); Keith Hawkins, *The Use of Legal Discretion*, *in* THE USES OF DISCRETION, *supra*, at 43 ("[I]t is clear from a large number of studies that assessments of moral character made by legal decision-makers are one of the most pervasive and persistent features in shaping the exercise of discretion."). Aaron Nielson offers an empirical and theoretical analysis in the context of administrative enforcement decisions in *How Agencies Choose Whether to Enforce the Law*, 93 NOTRE DAME L. REV. 1517, 1527–40 (2018) (considering survey data regarding agency use of enforcement discretion and concluding that limited resources and unforeseen circumstances sometimes require nonenforcement discretion, but that such instances should be rare given the risk of bias and loss of notice when public text of law does not match enforcement).

The problem is compounded by information asymmetries and the opacity of enforcement decision-making. Agencies often rely on internal guidelines or informal norms, making it difficult for the public to assess whether like cases are treated alike.[233] As Keith Hawkins has observed, discretion in enforcement is not rule-free, but it is often structured by organizational routines and expectations that operate below the surface of formal law.[234] These underlying constraints do not always track legal merit. They may reflect risk aversion, bureaucratic habit, or political calculation as much as they reflect justice.[235] The result is a system that may appear neutral in form but is deeply selective in practice. And that selectivity, once normalized, undermines the rule-of-law ideal that government should proceed by publicly known, generally applicable, and consistently enforced rules.

One example is the IRS, which seems as if it ought to be the paragon of neutrality, yet has faced allegations of targeting politically disfavored groups, such as Tea Party–affiliated not-for-profit organizations under the Obama administration.[236] Similarly, under the Trump administration, the Department of Education opened Title VI investigations into elite universities—including Harvard, Yale, and Columbia—in ways that critics charged were politically motivated.[237] These are not anomalies. In areas where discretion is broad and stakes are high, enforcement becomes a means for the exercise of power. That concern is magnified in rapidly evolving technological domains, where factual uncertainties give regulators still more room to maneuver.

---

[232] *See* Kent Barnett, *Against Administrative Judges*, 49 U.C. Davis L. Rev. 1643, 1650–83 (2016) (assessing the potential for partiality among administrative judges, which lack the independence protections of ALJs, and concluding that they present an unconstitutional appearance of impartiality); Kent Barnett, *Why Bias Challenges to Administrative Adjudication Should Succeed*, 81 Mo. L. Rev. 1023, 1031–43 (2016) (surveying Supreme Court administrative law decisions and concluding challenge to lack of independence of administrative judges follows the trend of precedent).

[233] *See* Alexander Nabavi-Noori, *Agency Control and Internally Binding Norms*, 131 Yale L.J. 1278, 1290–97 (2022) (describing how agencies rely on internal guidance documents and informal norms, which makes public oversight of consistency difficult).

[234] *See* Keith Hawkins, *The Use of Legal Discretion*, *supra* note 231, at 11–13, 39–41, 43 (considering the open-ended discretion of bureaucrats and cautioning that "much of what is often thought to be the free and flexible application of discretion by legal actors is in fact guided and constrained by rules to a considerable extent" and that such rules "tend not to be legal, but social and organizational," and concluding that "[i]t is the lack of fit between the legal expectations about how a decision should be made and how it is socially determined in practice which may give rise to accusations of arbitrariness or irrationality").

[235] *See* Sunstein, *supra* note 3, at 64–88 (explaining that official decisions often reflect risk aversion and political responses to fear).

[236] Juliet Eilperin, *Document details IRS scrutiny of Constitution-focused groups*, Wash. Post, May 13, 2013, at A8; Scott Wilson, *Justice Department, IRS Scandals Challenge Obama's Civil Liberties Credibility*, Wash. Post, May 15, 2013, at A7; Treas. Inspector Gen. for Tax Admin., Inappropriate Criteria Were Used to Identify Tax-Exempt Applications for Review (May 14, 2013), https://oversight.house.gov/wp-content/uploads/2013/05/201310053fr-revised-redacted-1.pdf [https://perma.cc/7KP6-5PFF]; *Hearing on Internal Revenue Service Targeting Conservative Groups: Hearing Before the H. Comm. on Ways & Means*, 113d Cong.

This dynamic calls into question the wisdom of granting new discretionary enforcement powers over technologies like artificial intelligence, user interfaces, or data-sharing contracts. Enforcement decisions, like regulations themselves, will be shaped by institutional pathologies. Legislation that assumes otherwise invites disappointment at best and injustice at worst.

2. *Principal-Agent Problems in Bureaucracy*

Even where outright bias and political motivations are absent, agencies and their staff suffer from internal misalignments of a more general sort. The classic principal-agent problem arises whenever the objectives of the agent (the agency or its employees) diverge from those of the principal (Congress or the public).[238] These problems are endemic in bureaucratic structures. As William Niskanen explained, bureaucrats cannot be assumed to operate purely in the public interest.[239] They are humans and, like employees of any other entity, will seek when possible to further their own interests, such as reduced workloads, career advancement, prestige, and job security, not just the public welfare. Because their performance is difficult to measure, and because

---

(2013) (testimony of J. Russell George, Treas. Inspector Gen. For Tax Admin.). *But see* Alan Rappeport, *In Vetting Political Groups, I.R.S. Crossed Party Lines*, N.Y. Times, Oct. 5, 2017, at A23; Phillip Hackney, *Should the IRS Never Target Taxpayers?—An Examination of the IRS Tea Party Affair*, 48 Val. U. L. Rev. 1 (2014) (reviewing available evidence and concluding IRS investigations to have followed appropriate procedures).

[237] *See* Emily Bazelon & Charles Homans, *The Battle over College Speech Will Outlive the Encampments*, N.Y. Times Mag. May 29, 2024, https://www.nytimes.com/2024/05/29/magazine/columbia-protests-free-speech.html [https://perma.cc/2F5W-NYW6]; *U.S. Department of Education Probes Cases of Antisemitism at Five Universities*, U.S. Dep't of Educ. (Feb. 3, 2025), https://www.ed.gov/about/news/press-release/us-department-of-education-probes-cases-of-antisemitism-five-universities [https://perma.cc/C37J-WNYZ]; U.S. House of Representatives, Staff Report on Antisemitism, at 25–26 (2024), https://www.speaker.gov/wp-content/uploads/2024/12/House-Antisemitism-Report.pdf [https://perma.cc/7GPM-FB6T]; *Trump Administration Scrutiny of Academic Institutions Stretches Beyond Elite Colleges*, NPR (July 24, 2025), https://www.npr.org/2025/07/24/g-s1-78918/trump-administration-scrutiny-of-academic-institutions-stretches-beyond-elite-colleges [https://perma.cc/2SFB-99A2].

[238] For general discussion of agency theory and the principal-agent problem, see Carol M. Kopp, *What Is Agency Theory?*, Investopedia (Sep. 30, 2024), https://www.investopedia.com/terms/a/agencytheory.asp [https://perma.cc/WM84-5JS4]; Thomas A. Lambert, How to Regulate: A Guide for Policymakers 91–134 (2017).

[239] *See* Niskanen, *supra* note 225, at 39 (discussing the impossibility of a neutral bureaucrat because of conflicting interests).

external accountability is weak, bureaucrats may make enforcement decisions that serve personal goals over statutory objectives.[240]

One consequence is risk-averse over-enforcement. Agency staff who fear adverse headlines or individual blame have an incentive to interpret rules expansively and demand more than the law requires, to reduce the risk of a public backlash for which they may personally suffer career consequences.[241] The same logic drives corporate compliance officers, who may recommend to their superiors compliance beyond what the law requires.[242] Better to over-comply than take any risk of a negative outcome for which they personally may suffer consequences. Such cautiousness, however, although personally beneficial, imposes real costs on innovation, which may be hindered by rules that were never really intended by Congress.[243]

Similarly, careerism may affect not just the level of agency enforcement, but how companies are selected for targeting. Enforcement decisions may be shaped not by policy merits but by the publicity they generate.[244] An ambitious regulator might pursue headline-making enforcement actions against popular targets, while neglecting more systemic but less visible harms. Worse, the so-called "revolving door" of employees leaving government service for industry and vice versa creates incentives to avoid taking action against firms where one hopes to land a job.[245]

A particularly vivid example is the U.S. Food and Drug Administration (FDA). The FDA is widely regarded as extraordinarily stringent in approving new drugs.[246] This strictness is not necessarily a reflection of superior risk

---

[240] *Id.* at 128.

[241] *Id.* at 221; *see also* Henry I. Miller, *The FDA: Challenges for a New Century, A Rough Road Ahead for Would-Be Reformers*, 11 N.Y.U. J.L. & Liberty 803, 821–27 (2025) (describing in the context of the FDA how unfamiliarity with the relevant technology can lead to overcaution).

[242] *See generally* Jennifer Pahlka, Recoding America (2023).

[243] *See* Susan Lorde Martin, *Compliance Officers: More Jobs, More Responsibility, More Liability*, 29 Notre Dame J.L. Ethics & Pub. Pol'y 169, 169–71 (2015) (explaining that fear of personal liability leads compliance officers to err on the side of caution, even when such over-compliance is costly to the firm); M. Kabir Hassan, Reza Houston, & M. Sydul Karim, *Courting Innovation: The Effects of Litigation Risk on Corporate Innovation*, 71 J. Corp. Fin. 102098, 1–22 (2021) (finding that litigation risk creates managerial myopia and deters investment in risky but innovative projects).

[244] *See* Margaret H. Lemos & Max Minzner, *For-Profit Public Enforcement*, 127 Harv. L. Rev. 853, 875–93 (2014) (explaining that public enforcers have reputational incentives to emphasize large, easily publicized penalties, and that agencies and individual lawyers may prioritize cases for the visibility they generate rather than for their policy merits).

[245] Michael A. Livermore & Richard L. Revesz, *Regulatory Review, Capture, and Agency Inaction*, 101 Geo. L.J. 1353, 1375 (2013) (noting the incentive for agency employees to develop relationships with regulated entities while in government to facilitate later private-sector employment). For overviews of the literature on the revolving-door concern, see James D. Cox & Randall S. Thomas, *Revolving Elites*, 107 Geo. L.J. 845, 853–59 (2019); Dal Bó, *supra* note 193, at 214–15.

[246] *See* Henry I. Miller & David R. Henderson, *The FDA's Risky Risk-Aversion*, 145 Pol'y Rev. 3, 3–27 (2007) (noting that the FDA is widely regarded as the "gold standard," meaning "they are the most stringent and risk-averse in the world," and describing the length and costs of U.S. drug approvals as the highest globally); Hans-Georg Eichler, Brigitte Bloechl-Daum,

assessment, but could instead reflect asymmetrical reputational incentives.[247] If the FDA approves a drug that later proves harmful, the agency is publicly blamed and its credibility damaged. But if the FDA delays or denies approval of a beneficial drug, the costs—measured in lost lives or prolonged suffering—are diffuse and largely invisible. Those denied a treatment they never received cannot be counted. The agency's staff face little reputational or career risk from excessive caution, but considerable risk from a visible mistake. As a result, regulators often favor caution even where it may harm the public more than it protects them.

Principal-agent dynamics ensure that such preferences will not always align with public interest. Sometimes, agency staff may be too cautious; sometimes, too aggressive. But they will not be neutral. That reality stands in contrast to the nirvana ideal often implicitly assumed by proponents of new tech regulation. When considering new laws to govern emergent technologies, it is not enough to ask what an ideal enforcement regime would do. One must ask who will do the enforcing, under what incentives, and with what information. Bias and misalignment are not unfortunate exceptions. They are features of real institutions. And any serious regulatory proposal must reckon with them.

*3.   Case Study: The Computer Fraud and Abuse Act*

The Computer Fraud and Abuse Act (CFAA)[248] stands as a cautionary example of what can go wrong when lawmakers respond to technological change with sweeping legislation. Enacted in 1986, the statute was originally conceived to address a narrow and novel threat: the rise of computer hacking.[249] But over time, its ambiguous language and expanding reach allowed prosecutors and civil litigants to repurpose the law in ways far removed from its original aim.[250] Only recently have courts restored a more limited interpretation to the Act.

---

Daniel Brasseur, Alasdair Breckenridge, Hubert Leufkens, June Raine, Tomas Salmonson, Christian K. Schneider & Guido Rasi, *The Risks of Risk Aversion in Drug Regulation*, 12 Nat. Rev. Drug Discovery 907, 907–16 (2013) (observing that regulators, including the FDA, have been criticized for excessive risk-aversion, requesting too much data and delaying approvals, which can deprive patients of potentially beneficial treatments).

[247] *See* Miller & Henderson, *supra* note 246, at 3–27 (explaining that FDA's extreme caution is driven by institutional incentives to avoid visible errors, not necessarily by superior risk assessment); *see also* Sunstein, *supra* note 3, at 41–45 (arguing that loss aversion drives regulatory behavior, with regulators more attuned to visible harms from approval than to hidden costs of delay).

[248] Computer Fraud and Abuse Act of 1986, Pub. L. No. 99-474, 100 Stat. 1213 (1986) (codified as amended in scattered sections of Titles 18, 15, and 12 of the U.S. Code).

[249] For a brief history of the CFAA, see Justin Precht, Note, *The Computer Fraud and Abuse Act or the Modern Criminal at Work: The Dangers of Facebook from Your Cubicle*, 82 U. Cin. L. Rev. 359, 359–62 (2014).

[250] *See* Samantha Jensen, Note, *Abusing the Computer Fraud and Abuse Act: Why Broad Interpretations of the CFAA Fail*, 36 Hamline L. Rev. 81, 81–95 (2013) (observing that the CFAA, originally narrow and aimed at hackers, was broadened by amendments and has been

Congress enacted the CFAA in 1986 as an amendment to earlier computer crime legislation in the Comprehensive Crime Control Act of 1984.[251] The 1986 Act was designed to target "the technologically sophisticated criminal who breaks into computerized data files,"[252] an archetype that would have brought to mind Cold War fears of Soviet espionage and sabotage. At the time, computers were unfamiliar and symbolically powerful. To access them without permission was to trespass on sensitive government or financial networks. The CFAA therefore criminalized "unauthorized access" to systems "operated for or on behalf of the Government of the United States"[253] and prohibited users from "exceeding authorized access" to obtain data from such systems.[254]

But the statute's limited scope did not last. In 1996, the Economic Espionage Act[255] dramatically expanded the CFAA's reach by extending its coverage to any "protected computer,"[256] defined broadly to include any device used in or affecting interstate or foreign commerce—essentially any internet-connected machine.[257] That same amendment also added to the CFAA's criminal liability provisions a private right of action, meaning that any person who "intentionally accesses a computer without authorization or exceeds authorized access" could now be subject not only to imprisonment for up to ten years,[258] but also to civil liability for any "damage or loss" caused.[259] As amended, the CFAA in effect created a federal trespass to chattel action for unauthorized access to any internet-connected data or computer system.[260]

Since the amendment, the legal architecture of the CFAA has remained the same, but society has continued to change around it. By the late 1990s, and certainly by the early 2000s, computers were no longer the province of elite operators, government contractors, or Cold War spies. They were in every office, on every desk, and, with the advent of smartphones, in every

---

increasingly invoked by employers against disloyal employees, extending the statute far beyond its original purpose).

[251] Comprehensive Crime Control Act of 1984, Pub. L. No. 98-473, 98 Stat. 1976 (1984) (codified as amended in scattered sections of Titles 18 and 28 of the U.S. Code).

[252] H.R. Rep. No. 99-612, at 3 (1986).

[253] 18 U.S.C. § 1030.

[254] Precht, *supra* note 249, at 360–62 (discussing the CFAA's history and its pre-amendment limitation to governmental computers).

[255] Economic Espionage Act of 1996, Pub. L. No. 104-294, 110 Stat. 3488 (1996) (codified as amended at 18 U.S.C. §§ 1831–39).

[256] 18 U.S.C. § 1030(e)(2).

[257] *See* Van Buren v. United States, 593 U.S. 374 (2021); George F. Leahy, *Keeping Gates Down: Further Narrowing the Computer Fraud and Abuse Act in the Wake of Van Buren*, 14 Wm. & Mary Bus. L. Rev. 215, 224–29 (2022) (discussing the circuit split that preceded the Court's decision in *Van Buren*).

[258] 18 U.S.C. § 1030(a)(2), (c)(1)(a), (g).

[259] *Id.* § 1030(g).

[260] *See* Riana Pfefferkorn, *Shooting the Messenger: Remediation of Disclosed Vulnerabilities as CFAA "Loss,"* 29 Rich. J.L. & Tech. 89 (2022); *Van Buren*, 593 U.S. at 397–408 (2021) (Thomas, J., dissenting) (analogizing CFAA to common law's historical protection against unauthorized use of one's property, including by exceeding the scope of the property owner's consent).

pocket. Employees used them routinely to send emails, check calendars, and access company files. Being the property of corporate employers, and not the employees themselves, employees' access to and use of those systems was often subject to usage policies, employment contracts, or terms-of-service agreements.[261] The problem was that the CFAA's language, under which "exceed[ing] authorized access" constitutes a violation,[262] could plausibly be read to criminalize and create civil liability for deviation from employer usage policies. An employee might violate the CFAA by something as simple as checking her email or logging into a social media account from her workplace computer.[263] Ordinary workplace conduct became grounds for civil or even criminal liability.

The CFAA's breadth created two distinct problems when applied to modern society in which computers are such commonplace tools.[264] First, the CFAA became a favored claim for private litigants to pursue business and workplace disputes under a federal anti-hacking statute designed for a different era.[265] Second, the statute handed prosecutors expansive discretion. With broad statutory language regarding unauthorized access, and computer systems now a ubiquitous technology, nearly any undesirable conduct involving computers could plausibly be framed as a CFAA violation.[266] This allowed enforcers to selectively invoke the statute against disfavored individuals and to attach significant criminal penalties to otherwise routine matters.[267]

The issue simmered for decades in the lower courts until, in 2021, the Supreme Court addressed the CFAA for the first time in *Van Buren v. United States*.[268] In that case, the defendant, a police officer, had accessed a license plate database for personal reasons in violation of his department's policy.[269] He had permission from the department to access the system, but only when used for legitimate purposes.[270] The government argued that although he was authorized to use the database, he had violated the CFAA by using the database in a manner contrary to his department's policy.[271] A divided Supreme Court disagreed.[272] The Court held that a person "exceeds authorized access" only by accessing information that she is not entitled to obtain, not when using accessible information for an improper purpose.[273]

---

[261] *See* Orin S. Kerr, *Cybercrime's Scope: Interpreting "Access" and "Authorization" in Computer Misuse Statutes*, 78 N.Y.U. L. Rev. 1596, 1637–40 (2003) (explaining that courts have sometimes treated violations of contractual terms such as workplace computer-use policies as exceeding authorization under the CFAA).

[262] 18 U.S.C. § 1030(a)(2).

[263] *See* Jensen, *supra* note 250, at 95–109 (discussing the difficult statutory interpretation questions posed by the text of CFAA and the practical problems presented by its potential breadth).

[264] *See* Pfefferkorn, supra note 260, at 91 (noting CFAA's potential application to behavior "far afield from the law's core anti-hacking purpose"); Orin S. Kerr, *Focusing the CFAA in Van Buren*, 2021 Sup. Ct. Rev. (2022) (retrospective discussion of the Supreme Court's decision in *Van Buren*).

[265] *See* Patricia L. Bellia, *Defending Cyberproperty*, 79 N.Y.U. L. Rev. 2164, 2232–41 (2004) (explaining that companies invoked the CFAA in civil suits to restrict unwanted uses of their systems, extending the statute beyond hacking into business disputes).

[266] *See* Orin S. Kerr, *Vagueness Challenges to the Computer Fraud and Abuse Act*, 94 Minn. L. Rev. 1561, 1561–65 (2010) (explaining that amendments and broad language have expanded the CFAA so far that nearly any computer-related misconduct could be captured within its scope); Andrew T. Hernacki, *A Vague Law in a Smartphone World: Limiting the Scope of Unauthorized Access Under the Computer Fraud and Abuse Act*, 61 Am. U. L. Rev. 1543, 1543–48 (2012) (showing how litigants have attempted to stretch the CFAA to cover conduct far afield from the traditional computer hacking, such as mobile app data collection).

[267] *See* Kerr, *supra* note 266, at 1562–63 (warning that vague CFAA provisions would give prosecutors such wide discretion so as to fail to provide due process of law).

[268] 593 U.S. 374 (2021).

In support of its conclusion, the *Van Buren* Court emphasized that the government's interpretation would "criminalize everything from embellishing an online-dating profile to using a pseudonym on Facebook," all based on breaches of terms of use.[274] That result, it concluded,[275] would render the CFAA dangerously vague and overbroad—precisely the concerns that had animated academic and judicial criticism for years.[276]

Yet the case was not an easy decision. The basic question had split the federal circuit courts for decades[277] and ultimately divided the *Van Buren* Court 6-3.[278] The majority was partly seeking to avoid criminalizing such a broad swath of ordinary computer activity,[279] and the dissenting Justices were concerned that by contemplating the CFAA's consequences as applied to modern society's routine computer usage, the majority improperly assumed that a Cold War-era Congress of the 1980s was aware of how computers would be used in 2021.[280] The case thus illustrates another difficulty of lawmaking

---

[269] *Id.* at 378.

[270] *Id.*

[271] *Id.* at 380.

[272] For detailed analysis of the Court's *Van Buren* decision and the confusion and division of authority that preceded it, see generally Kerr, *supra* note 264.

[273] *Van Buren*, 593 U.S. at 396.

[274] *Id.* at 394 (quoting Brief of Orin Kerr as Amicus Curiae 10–11).

[275] *Id.* at 394–96 (reasoning that "[i]f the 'exceeds authorized access' clause criminalizes every violation of a computer-use policy, then millions of otherwise law-abiding citizens are criminals").

[276] *See, e.g.*, Kerr, *supra* note 266, at 1562 ("The CFAA has become so broad, and computers so common, that expansive or uncertain interpretations of unauthorized access will render it unconstitutional. Such interpretations would either provide insufficient notice of what is prohibited or fail to provide guidelines for law enforcement in violation of the constitutional requirement of Due Process of the law."); Patricia L. Bellia, *A Code-Based Approach to Unauthorized Access Under the Computer Fraud and Abuse Act*, 84 Geo. Wash. L. Rev. 1442, 1470–74 (2016) (arguing that broad interpretations of the CFAA's "unauthorized access" provisions risk unconstitutional vagueness and urging a narrower approach that would turn on the user bypassing a "code-based" barrier); Kerr, *supra* note 261, at 1646–48 (suggesting that unauthorized access be construed narrowly to include only cases involving circumvention of access controls).

[277] *See* Precht, *supra* note 249, at 362–63 (describing the longstanding division among federal circuits).

[278] *See* 593 U.S. 374 (2021).

[279] *Id.* at 376 (noting the "breathtaking amount of commonplace computer activity" that would fall within the CFAA if interpreted as suggested by the Government).

[280] *Id.* at 407 (Thomas, J., dissenting) (criticizing "majority's reliance on modern-day uses of computers to determine what was plausible in the 1980s" to Congress when it enacted the statute).

in the face of technological change: even a well-intentioned law designed to combat what at the time was a narrow problem may come to (poorly) govern a much broader range of human conduct than was ever intended as society's use of that technology shifts.

### E.  *Ossification and Technological Change*

Legal ossification poses still another problem to lawmakers in an age of technological change. As discussed in the prior section,[281] even narrow, well-intentioned laws can outlive their usefulness and come to distort entire domains of conduct if the role the governed technology plays in society changes over time. That is a type of legal ossification, but the problem goes beyond merely allowing outdated rules to persist. It may even prevent legal systems from accomplishing lawmaking in the first place. Put differently, the problem is not only that legislative and regulatory law may be slow to catch up, but also that it may be structurally incapable of doing so. This section discusses the problems that legal ossification poses in the context of rapidly changing technology, especially for laws structured as rules rather than standards.

#### 1. *The Pacing Problem of Innovation and Regulation*

In administrative law, ossification refers to the increasing procedural and institutional rigidity that impedes the timely development of new law.[282] As Thomas McGarity described in his foundational account, the rulemaking process has become "increasingly rigid and burdensome," weighed down by layers of analytical mandates, judicial doctrines, and internal agency review that frustrate efforts to regulate even modest issues.[283] These constraints can transform a process once envisioned as nimble and participatory into one marked by delay, overcautious drafting, and institutional inertia. The dominant view among administrative law scholars is that these delays are not incidental, but structural, and symptomatic of a process that has become "so heavily laden with additional procedures, analytical requirements, and external review mechanisms"[284] that its comparative advantage over case-by-case adjudication has eroded.[285] Even routine agency rules now require substantial time and legal resources, as officials prepare not only for public comment but for inevitable legal challenges from courts, overseers, and stakeholders alike.[286]

---

[281] *See supra* Part IV.D.3.
[282] *See* McGarity, *supra* note 17, at 1385–1405 (explaining how added procedural mandates and judicial review have made rulemaking slow, rigid, and litigation-prone).
[283] *Id.* at 1385.
[284] *Id.* at 1385–87 (describing how rulemaking, once praised as efficient and flexible, has "become increasingly rigid and burdensome").
[285] *Id.* (rulemaking processes' "superiority to case-by-case adjudication is not as apparent now as it was before it came into heavy use"); *see also* Richard J. Pierce, Jr., *Rulemaking Ossification Is Real: A Response to Testing the Ossification Thesis*, 80 Geo. Wash. L. Rev. 1493, 1493–95 (2012) (summarizing the ossification thesis and efforts to measure it empirically).

The causes of ossification are many. The cumulative layering of analytical mandates, cost-benefit requirements, environmental assessments, and executive oversight has made informal rulemaking functionally indistinct from formal adjudication in its complexity. Agency lawyers, operating in what Thomas McGarity has called the "team model," play a central role in this process by working to ensure that proposed rules are not just substantively defensible but procedurally impeccable.[287] In this context, lawyers are empowered to "veto aspects of proposed rules that they find to be unlawful" or even to reshape them "to fit their own policy preferences or what they deem to be the policy preferences of the reviewing judges."[288] The result is often paralysis by anticipation—rules written not just for public comment but for imagined litigation.

This phenomenon is troubling enough in stable domains. In dynamic domains, such as AI, digital commerce, or cybersecurity, it is catastrophic. In these contexts, technology changes not just the application of the law but the very contours of the problem to which law might respond. A regulation that was narrowly tailored to one form of digital deception, for example, may be useless or even counterproductive by the time it takes effect.[289] Moreover, each law passed to curb some new undesirable practice may foreclose entirely different innovations before they are even born.[290] The result is a legal landscape simultaneously rigid and misaligned—incapable of regulating the present and incapable of anticipating the future. Technological change thus exacerbates ossification's most damaging effects. The problem is not merely that rulemaking is slow, but that it assumes a world that no longer exists by the time the rule arrives. The gap is visible in fields as diverse as biotechnology and copyright law. In biotechnology, legislative responses have struggled to keep pace with rapid technological advances in DNA processing capacity, leaving regulators either to apply outdated statutes or to improvise through guidance documents.[291] In copyright, Congress's attempt at technology-neutral drafting in the 1976 Act was undermined as new communications technologies such as home video recorders and digital streaming produced disputes unforeseen by lawmakers.[292] This general obstacle to lawmaking is known as the "pacing problem," in which law's slowness collides with the velocity of innovation.[293] The resulting bind is that lawmakers are forced to choose between "reckless action (regulation without sufficient facts)" or "paralysis (doing nothing at all)."[294]

---

[286] *See* Thomas O. McGarity, *The Role of Government Attorneys in Regulatory Agency Rulemaking*, 61 L. & Contemp. Probs. 19, 22 (1998) (discussing agency attorneys' need to anticipate legal challenges of their work).

[287] *See id.* at 20–21 (discussing the team model).

[288] *Id.* at 20.

[289] *See* Dickinson, *The Patterns of Digital Deception*, *supra* note 143, at 2486–97 (making this point regarding technology generally and dark patterns in particular).

[290] *See id.* at 2486–97 (arguing that technology-specific, rule-based regulations on multipurpose tools like A/B testing, dark patterns, and generative AI are prone to both over- and under-inclusiveness, risking harm to innovation while failing to capture evolving deceptive practices); *see, e.g.*, *supra* note 187 (collecting sources assessing the GDPR's effect on app innovation in the EU).

[291] *See* Gary E. Marchant, *The Growing Gap Between Emerging Technologies and the Law*, *in* The Growing Gap Between Emerging Technologies and Legal Ethical Oversight: The Pacing Problem 19, 19–33 (Gary E. Marchant, Braden R. Allenby & Joseph R. Herkert eds., 2011) (describing how law lags far behind emerging biotechnologies).

### 2. *Rules and Standards in Technology Regulation*

Ossification is especially problematic when lawmakers adopt rule-based rather than standard-based statutes and regulations.[295] Rules employ a highly definite test of applicability. Determining whether a rule applies requires resolution only of simple questions of fact[296]—for example, whether a defendant used a specified prohibited technology or dataset when evaluating applications for credit or home loans.[297] Definite rules of this sort are often preferable to flexible standards because they make it possible for regulated parties to organize their affairs with greater certainty and at lower cost.[298] For example, most drivers prefer posted speed limits to a prohibition on driving at an "unreasonable rate of speed."[299] Yet rules, as Louis Kaplow famously explains, "are more costly than standards to promulgate because rules involve advance determinations of the law's content, whereas standards are more costly for legal advisors to predict or enforcement authorities to apply because they require later determinations of the law's content."[300] Rules' ex ante clarity is thus their greatest strength, but it becomes a liability in contexts of rapid change, where the costliness of the rulemaking process inhibits adaptation.

---

[292] *See* Brad A. Greenberg, *Rethinking Technology Neutrality*, 100 Minn. L. Rev. 1495, 1495–1501 (2016) (arguing that even the Copyright Act of 1976, though drafted with technology-neutral defaults, quickly ossified with the development of new communications technologies).

[293] For further discussion regarding the pacing problem, see Hagemann et al., *supra* note 197, at 58–60; Marchant, *supra* note 291, at 22–23; Wendell Wallach, A Dangerous Master 295–400 (2015).

[294] *See* Mark D. Fenwick, Wulf A. Kaal & Erik P.M. Vermeulen, *Regulation Tomorrow: What Happens When Technology Is Faster than the Law?*, 6 Am. U. Bus. L. Rev. 562, 568–73 (2017) (describing how the pacing problem, in which the accelerating speed of technological innovation outstrips the slower pace of legislative and regulatory processes, forces policymakers into a choice between premature, underinformed action and regulatory inaction).

[295] For detailed discussion of the tradeoffs between rules and standards, see Louis Kaplow, *Rules Versus Standards: An Economic Analysis*, 42 Duke L.J. 557 (1992); Pierre Schlag, *Rules and Standards*, 33 UCLA L. Rev. 379 (1985); Gideon Parchomovsky & Alex Stein, *Catalogs*, 115 Colum. L. Rev. 165, 172–81 (2015).

[296] *See* Henry M. Hart, Jr. & Albert M. Sacks, The Legal Process 138–40 (Foundation Press, Inc. 1994) (1958).

[297] See, for example, Equal Credit Opportunity Act (ECOA), Pub. L. No. 93-495, 88 Stat. 1521 (1974) (codified as amended at 15 U.S.C. §§ 1691–1691f), § 1691 of which requires prospective lenders to provide "a statement of reasons for [any adverse action]." *Consumer Financial Protection Circular 2022–03: Adverse Action Notification Requirements in Connection with Credit Decisions Based on Complex Algorithms*, Consumer Fin. Prot. Bureau (May 26, 2022), https://www.consumerfinance.gov/compliance/circulars/circular-2022-03-adverse-action-notification-requirements-in-connection-with-credit-decisions-based-on-complex-algorithms/ [https://perma.cc/BF2Q-SFCU] (interpreting the adverse action notice requirements to preclude the use of black-box AI models that cannot produce specific reasons for credit denial); *see also* Cal. Civ. Code § 1785.13(a)(2) (West 2025) (barring credit reporting agencies from including certain information more than seven years old); S.B. 605, 83d Legis. Assemb., Reg. Sess. (Or. 2025) (prohibiting reporting of medical debt to credit reporting agencies).

[298] *See* Kaplow, *supra* note 295, at 559–68 (explaining that rules provide greater ex ante certainty and allow regulated parties to structure their conduct at lower cost).

Standards, by contrast, delay specificity. They articulate general principles (for example, reasonableness, good faith, or fairness) and leave their application to future decisionmakers who can consider the facts as they unfold.[301] This design makes standards better suited for environments of uncertainty, where context matters more than categorical prediction. In practice, this means that when the facts change faster than Congress or agencies can react, courts interpreting broadly worded statutes or common-law doctrines will typically prove more responsive than legislatures.[302]

\* \* \*

This Part has detailed multiple pathologies[303] that afflict statutory and regulatory lawmaking. The point of doing so is not to condemn positive lawmaking or legal change, both of which are necessary parts of our legal system, but to explain why the common law and similarly broad "common law statutes" are often better positioned to respond to technological change than are more particular enactments.

Legislators may feel compelled to "do something" to confront each new abuse of technology. Yet in doing so, they do not make the mistake merely of repetition—of "double banning" practices already prohibited by general law. The point of this Part has been to show how, by enacting law specific to new technologies, legislators are also likely to do positive harm. By even attempting prophylactically to prohibit or limit the use of certain technologies, lawmakers, who are necessarily acting on incomplete information, may succumb to regulatory capture, introduce opportunities for biased and self-serving enforcement by agency bureaucrats, and hinder the development of innovative technologies, all in pursuit of a nirvana that cannot exist.

---

[299] *See* Thomas A. Lambert, How to Regulate: A Guide for Policymakers 101 (2017) (discussing Montana's experiment with a common-law reasonableness standard for regulating speeding from 1995 to 1998).

[300] Kaplow, *supra* note 295, at 562–63.

[301] Hart & Sacks, *supra* note 296, at 140 (defining standards as legal directives that require for their application, beyond findings of fact, "a qualitative appraisal of those happenings in terms of their probable consequences, moral justification, or other aspect of general human experience").

[302] For a discussion of the law's "response modes" to new technologies, see generally Urs Gasser, *Recoding Privacy Law: Reflections on the Future Relationship Among Law, Technology, and Privacy*, 130 Harv. L. Rev. F. 61 (2016) (taking privacy law as an example and observing that courts' adaptation of existing common-law and statutory frameworks to incorporate new technologies is the legal system's default solution).

[303] *See supra* Part IV.B–E.

## V. Conclusion: Directions for a Law-Proof Future

The human impulse to legislate in response to technological change is entirely natural, but it is very often misguided. Through the lenses of history, economics, and political science, this Article has urged that the default response to innovation—new statutes, new agencies, and new prohibitions of limitless variety—should be resisted, not merely out of caution but out of principle. The history of technological progress is not a record of disasters narrowly averted by preemptive legislation. It is, instead, a chronicle of steady social change and legal adaptation—often slow, sometimes in spurts, but effective. The challenge for lawmakers is to channel that adaptive capacity without undermining it in the name of precise control or Pyrrhic political victories. This Article thus concludes by stepping beyond critique to offer some direction for a law-proofed future. If particularized lawmaking is doomed to failure, how then should emerging technology be governed? Four principles offer a starting point.

*Default to Generality*—First, even in the highest of high-tech industries, lawmakers should default to generality. The genius of the common law lies in its resistance to specificity. Property, tort, and contract are not obsolete doctrines awaiting replacement by algorithmically applied chapters and subchapters of the U.S. Code. They are simple frameworks, yet capacious enough to accommodate all manner of change because they are designed around relationships, not technologies.

The desire for granularity—fresh statutes for social media, generative AI, and targeted advertising—is understandable but misplaced. The law can best provide stability, certainty, and fairness only if it is allowed to apply generally, to all persons, in all contexts,[304] not piecemeal according to who can exert the greatest sway among lawmakers.[305] This is not a defense of the status quo for its own sake. It is a recognition that specificity and durability are in balance. To hard code today's concerns into tomorrow's legal environment is to invite obsolescence and inflexibility at scale.

In a dynamic world, general rules endure. To allow them to operate, we must resist the allure of bespoke laws tailored to current technologies.[306] If a

---

[304] *See* Epstein, *supra* note 6, at 253–56 (articulating the purpose of a legal system as to "establish with reasonable certainty the boundaries in which individual decisions can take place" by defining "original property holdings," providing a "law of contracts," and "protecti[ng] . . . persons and property . . . from aggression").

[305] *See supra* Part IV.C (discussing the risk of regulatory capture especially by powerful tech-industry incumbents).

[306] *See* Epstein, *supra* note 7, at ix–xiv (arguing that "permanence and stability are the cardinal virtues of the legal rules that make private innovation and public progress possible" and

new tool causes harm, the question is not "What new law shall we make?" but "What general principles apply here?" Often, the right answer will be found not in congressional novelty but in common-law continuity.

*Institutional Humility*—A second key principle is institutional humility—a recognition that legislatures, regulators, and experts alike operate under conditions of radical epistemic constraint. We do not know the full effects of new technologies. But more importantly, we *cannot* know them in advance. The knowledge problem, as articulated by Friedrick Hayek, is not merely about access to facts. It is about the structure of knowledge itself—dispersed, tacit, contextual, and emergent.[307]

Our lack of knowledge calls for restraint. The public law of the future must respect its own cognitive limits. When regulators act as if they can anticipate downstream consequences, they risk freezing innovation based on upstream fears. The history of moral panics around media technologies—printing, novels, radio, comic books, television, and social media—demonstrates a pattern of fear and a recurring failure of foresight.[308] Policymakers imagined catastrophe; what followed was adaptation and social betterment not because of, but in spite of, their actions.[309]

Institutional humility requires understanding that attempts to engineer outcomes through top-down legislation often fail, not because legislators' intentions are bad, but because the available information is insufficient. It requires recognizing that the costs of mistaken precaution can outweigh the risks of cautious permission.

*Let Courts Do Their Work*—Third, the legal system must be allowed the space and time to adapt to change incrementally, via the adaptive mechanism it already contains—the judiciary. Courts, which resolve actual disputes with known facts between opposing parties, each of whom has an interest in persuading the court of its position, are well-situated in contrast with forward-looking legislatures to gradually integrate new facts into existing law without reengineering the law's fundamental architecture. They do so not by inventing new rules for each new technology, but by applying existing ones to novel contexts. The doctrines of trespass, negligence, products liability, fiduciary duty, misrepresentation, and fraudulent inducement of contract are not frozen.

---

that "a legal regime that embraces private property and freedom of contract is the only one that in practice can offer permanence and stability").

[307] *See supra* Part IV.B; Hayek, *supra* note 14, at 522–24 (discussing the paramount importance of "the knowledge of the particular circumstances of time and place," which "by its nature cannot enter into statistics and therefore cannot be conveyed to any central authority in statistical form" and concluding therefore that discretion must be left to the "man on the spot" to make decisions).

[308] *See generally* Cohen, *supra* note 8, at vi–xliv (discussing recurring social phenomenon of moral panics and their mildness in retrospect).

[309] *See supra* Part II (discussing the examples of the printing press, "brain-rot" media, and the elevator).

They are frameworks designed to accommodate change and all manner of social wrongs, both those known and those still unknown.

Richard Epstein has called this the "static conception" of the common law, and he is right to emphasize the law's continuity both as a descriptive and prescriptive matter.[310] But the term can also be misleading. Static does not mean stagnant. It means stable enough to allow society to change without requiring the law to constantly remake itself.[311] It is the legislature more often than the judiciary that imagines that each new tool demands a new rule. Indeed, the power of the judiciary lies precisely in its capacity to adjudicate the unknown without attempting to control it. Litigation is retrospective, fact-sensitive, and limited in scope. Judicial decision-making, not prescriptive legislation, should be the leading tool of governance in a world of experimentation.

*Let Innovation Proceed by Default*—Finally, we should resist the increasingly common view that innovation must earn its legality in advance. The better default is the opposite: that innovation proceeds unless and until it demonstrably causes harm. This is the foundational premise of what Adam Thierer has dubbed "permissionless innovation."[312] It is not an anti-law posture; it is a jurisprudential allocation of burdens. It says that new ideas and new technologies are presumptively allowed, that is, that the legal system will respond to real injuries when they occur but will not presume them into existence beforehand.

The permissionless approach does not imply regulatory abdication. It implies sequence. It allows society to discover the benefits of experimentation before moving to restrict it. The alternative is to require innovators to secure ex ante approval—a process that will inevitably entrench incumbents, delay entry, and suppress small-scale experimentation. This posture is especially important because the benefits of innovation are usually unorganized, distributed, and slow to surface, while the risks are often immediate, visible, and exaggerated. As Thierer has emphasized, the cost of regulatory precaution is not just foregone efficiency. It is also lost possibility.[313] "When public policy is shaped by precautionary principle reasoning," he writes, "it poses a serious threat to technological progress, economic entrepreneurialism, social adaptation, and long-run prosperity."[314]

---

[310] Epstein, *supra* note 6, at 253–76.

[311] *See supra* Part III; Epstein, *supra* note 6, at 253–56 (observing that "[t]here are some rules for which it is possible to observe . . . social depreciation," but that "in case after case, the proper result can and should be reached without taking into account any alleged dynamic element of the common law" and that although "[s]ocial circumstances continually change . . . it is wrong to suppose that the substantive principles of the legal system should change in response").

[312] THIERER, *supra* note 5, at vii–x, 1–11 (coining the phrase "permissionless innovation" and arguing that the principle should prevail over precautionary restrictions, with regulation justified only by clear evidence of harm).

[313] *Id.* at 32–34 (emphasizing that the costs of precautionary regulation include not only lost efficiency but also the loss of unforeseeable possibilities, since innovation foreclosed can never be realized).

[314] *Id.* at 16–18.

The lesson is not that innovation should go unregulated. It is that legal restrictions should follow innovation, not precede it. If injury occurs, the law can respond through remedies—through tort, through contract, through consumer protection, or through public enforcement. But those responses should come after the harm, not before the idea. The case for permissionless innovation is not that technology will never produce harm. It is that human judgment—legal, ethical, commercial, and social—is more effectively exercised once the shape of the problem is known. Governing the unknown through preemptive statutes requires speculative fear, and speculative fear is a poor substitute for reasoned response. The general rule should be simple: innovation is allowed. If some new technology violates a duty, deceives a consumer, invades a right, or causes some other compensable harm, the law will hold the innovator accountable. But the law should not require permission in advance.

*   *   *

Innovation, then, need not be lawless, and the law need not be hostile to innovation. A system committed to both liberty and progress must place its trust not in anticipatory design, but in adaptive response to human innovation. This is the genius of general law: it allows society to experiment boldly while preserving the stable principles needed to resolve conflict and protect rights. Our task is not to anticipate the future by code or statute, but to preserve the conditions under which law and society can meet that future—one case, one controversy, and one insight at a time, building over years a body of rules that is both durable and capable of growth. In that way, the law remains a steady companion to change rather than a brittle obstacle to it.